\documentclass[fleqn,usenatbib,useAMS]{mnras}

\usepackage{subfigure}
\usepackage{fancyhdr}
\usepackage{booktabs}
\usepackage{graphicx}
\usepackage{threeparttable}
\usepackage{adjustbox}
\usepackage{csquotes}

\usepackage{graphicx}		
\usepackage{amsmath}		
\usepackage{amssymb}		
\usepackage{multicol}        	
\usepackage{bm}			
\usepackage{pdflscape}		

\usepackage[T1]{fontenc}
\usepackage{ae,aecompl}

\usepackage{newtxtext,newtxmath}

\newcommand{\ignore}[1]{}

 \title[]{White dwarfs identified in LAMOST Data Release 5}
 \author[Jincheng Guo et al.]{Jincheng Guo$^{1,2,3,4}$\thanks{E-mail: andrewbooksatnaoc@gmail.com/jincheng.guo@ucl.ac.uk} \thanks{LAMOST fellow}, 
 Jingkun Zhao$^{5,6}$, 
 Huawei Zhang$^{3,4}$, 
 Jiajun Zhang$^{5,6}$, 
 Yu Bai$^{5}$, \newauthor
 Nikolay Walters$^{1}$,
 Yong Yang$^{5,6}$,
 and Jifeng Liu$^{5,6}$\\
 $^{1}$Department of Physics and Astronomy, University College London, London WC1E 6BT\\
 $^{2}$Beijing Planetarium, Xizhimenwai Road, Beijing 100044, China\\
$^{3}$Department of Astronomy, Peking University, Beijing 100871, China\\
$^{4}$Kavli Institute for Astronomy and Astrophysics, Peking University, Beijing 100871, P. R. China\\
$^{5}$Key Laboratory of Optical Astronomy, National Astronomical Observatories, Chinese Academy of Sciences, Beijing 100012, China\\
$^{6}$University of Chinese Academy of Sciences, Beijing 100049, People's Republic of China }

\begin{document}
 \date{}
 \pagerange{\pageref{firstpage}--\pageref{lastpage}} \pubyear{2019}
 \maketitle
 \label{firstpage}

\begin{abstract}
In this paper, we report white dwarfs identified in the 5$^{\rm th}$ Data Release of the Large Area Multi-Object fibre Spectroscopic Telescope, including spectral types of DA, DB, DC, DZ, and so on. There are 2\,625 DA spectra of 2\,281 DA stars, 182 DB spectra of 166 DB stars, 62 DC spectra of 58 DC stars, 36 DZ spectra of 33 DZ stars and many other types identified, in addition to our previous paper (Data Release 2). Among those sources, 393 DA stars and 46 DB stars are new identifications after cross-matching with the literature. In order to select DA candidates, we use the classification result from the LAMOST pipeline, colour-colour cut method and a random forest machine learning method. For DBs, since there is no template for DB in the pipeline model, a random forest machine learning method is chosen to select candidates. All the WD candidates have been visually checked individually. The parameters of effective temperature, surface gravity, mass, and cooling age have been estimated for relatively high signal-to-noise ratio DAs and DBs. The peaks of the DA and DB mass distributions are found to be around  0.62\,M$_{\odot}$ and  0.65\,M$_{\odot}$, respectively. Finally, the data and method we used to select white dwarf candidates for the second phase of LAMOST survey are also addressed in this paper.

\end{abstract}

\begin{keywords}
stars: white dwarfs; stars: fundamental parameters; Astronomical data bases: surveys-catalogues
\end{keywords}

 \section{Introduction}
The great majority of main-sequence stars are in the mass range of 0.07\,M$_{\odot}$ to 10\,M$_{\odot}$, which is also the mass range of white dwarf (WD) progenitors \citep{Doherty2015}. According to the literature \citep{Fontaine2001,Heger2003}, up to 97\% of all stars in our Galaxy will eventually evolve to WDs. Based on spectroscopic features, WDs mainly consist of DAs and DBs. They are the most explored types of single WDs, one reason is that they account for 90\% of WDs. The specific number of this fraction recently was found vary with the effective temperature \citep{Rolland2018,Cunningham2020}, based on $Gaia$ spectroscopic sample, after significant fraction of cool WDs were discovered \citep{Gentile2019}. Another reason is that they are objects of great importance, providing crucial information in various fields. For instance, the simple cooling mechanism of WDs makes it easier to obtain relatively accurate ages. Therefore, for research on the age of Galactic stellar halo \citep{Guo2016,Kilic2019,Guo2019}, WDs are important age estimation tools after their physics are well-understood. By studying mass distribution \citep{Kepler2007,Holberg2016,Hollands2018} and luminosity functions \citep{Harris2006,Munn2017,Lam2019} of WDs, multiple astrophysical processes of scientific importance can be learnt, in particular, the initial mass function and binary interactions. Meanwhile, WDs can serve as accurate records of star formation and reveal the evolution history of the Milky Way \citep{Krzesinski2009,Rowell2013}. Additionally, for stars that will eventually evolve to WDs, studies of their initial-final mass relation depend on both single WDs in clusters \citep{Catalan2008,Kalirai2008,Kalirai2009} and WDs in binaries \citep{Catalan2008b,Zhao2012}, These reseaches will help to provide important information on the evolution of our Galaxy \citep{Kilic2017}.

The goal to establish a large WD database has been pursued for more than three decades. \cite{McCook1987} spectroscopically identified 1\,279 WDs that has been updated to 2\,249 entries by \cite{McCook1999}. The total number of WDs has increased greatly, with the development of large surveys, e.g. Palomar-Green \citep{Green1986} and Sloan Digital Sky Survey (\citealt{York2000}, SDSS) particularly \citep{Kleinman2004,Eisenstein2006}. By using the data release (DR) 7 \citep{Abazajian2009}, \cite{Kleinman2013} spectroscopically identified about 20\,000 WDs. Recently, \cite{Kepler2015} made $\sim$ 9\,000 new identifications of WDs from SDSS DR10 \citep{Ahn2014}. More recently, around 6\,000 and 20\,000 WDs were identified from SDSS DR12 and DR14, respectively \citep{Kepler2016,Kepler2019}.

The Large Area Multi-Object fibre Spectroscopic Telescope (LAMOST) pilot survey started in 2012, and the first phase was completed five years later. After a transition from 2017 September to 2018 June, the survey began its second phase. There has been medium resolution (resolving power $R\sim$ 7500) spectroscopy carried out during bright nights in the second phase, together with lower resolution observations during dark nights ($R\sim$ 1800). In the five years of the first phase, several studies on LAMOST WDs have been conducted. \cite{Zhao2013} identified 70 DAs from the LAMOST pilot survey. \cite{Zhang2013} presented a catalog of 230 DAs by fitting Sersic profiles to Balmer lines of spectra. Combining the spectral type results from LAMOST pipeline, the Balmer line equivalent width measurements, and the colour-colour cut method, \cite{Guo2015b} identified 1056 DAs, 34 DBs, 276 white dwarf main sequence (WDMS) binaries and other spectral types of WDs in LAMOST DR2. \cite{Gentile2015} also reported the discovery of 253 new WDs in LAMOST DR3. By exploiting a well characterised magnitude limited DA star sample selected from LAMOST Galactic anticeter, space density, formation rate, luminosity and mass functions of DA WDs were studied \citep{Rebassa2015}. Recently, a catalog of 876 WDMS binaries identified in LAMOST DR5 are presented by \citet{Ren2018}.

Apart from spectroscopic surveys, the studies of WDs has opened a new window since the $Gaia$ data released \citep{Gaia2016}. Based on $Gaia$ data, local complete WD samples have been built  separately for 20\,pc \citep{Hollands2018} and 40\,pc \citep{Tremblay2020}. There are other much larger but not volume complete WD sample discovered in $Gaia$ DR2 \citep{Gaia2018}. \cite{Jimenez2018} presented a catalogue of 73\,221 WD candidates extracted from  DR2. More recently, more than 260\,000 high-confidence WD candidates were discovered in DR2, as well \citep{Gentile2019}. Evidence for merged WDs has also been revealed \citep{Kilic2018}. By adopting accurate positions, proper motions and parallaxes, detailed studies on WD kinematics have been conducted by several research groups \citep{Bovy2017,Gaia2018b,Rowell2019,Torres2019}.

In this work, we present a catalogue of WDs discovered in LAMOST DR5\footnote{Spectra available at \url{http://dr5.lamost.org/}}, as well as their physical parameter estimation. In Section 2, we present the selection methods of DAs and DBs. The estimation of the parameters for DAs and DBs (effective temperature, surface gravity, mass, and cooling age) is described in Section 3. We discuss our white dwarf target selection for the second phase of LAMOST in Section 4. Finally, we present our summary in Section 5.\\

\section{Candidates selection}

\subsection{LAMOST observation}

LAMOST is a 4-m reflecting Schmidt telescope, equipped with a multi-object spectrograph that has a 20\,deg$^{2}$ field of view and 4\,000 fibres. The LAMOST project began its one-year pilot study in 2012, followed by a five-year first phase survey. There is only low resolution spectroscopy in the first phase. The spectral resolving power is $R \sim 1800$, covering a wavelength range $3800-9000$ \,\AA. With typical exposure times of 1.5\,h, the limiting magnitude can reach 20.5\,mag \citep{Cui2012}. The first phase survey is composed of two major parts \citep{Zhao2012}. The first part is the LAMOST Experiment for Galactic Understanding and Exploration (LEGUE) survey, focused on understanding the structure and evolution of the Milky Way \citep{Deng2012}. The second part is the LAMOST Extra-Galactic Survey of galaxies (LEGAS), whose purpose is to study the large-scale structure of the Universe. The LEGUE survey consists of three smaller surveys, where each is selected for distinct purposes \citep{Chen2012,Carlin2012,Yuan2015}. The targets are mainly chosen from the Galactic disc, spheroid, and anti-center. Although LAMOST is a spectroscopic survey only, there are also photometric data provided by different astronomers from various photometric catalogs, e.g. XSTPS-GAC \citep[Xuyi Schmidt Telescope Photometric Survey of the Galactic Anti-center]{LiuXW2014,Guo2018}, 2MASS, SDSS, Kepler, NVSS etc. More than 50\% of the input catalog entries at least have $g$, $r$, $i$ magnitudes, mainly from XSTPS-GAC and SDSS. In LAMOST DR5, there are almost 10 million spectra (details in Table \ref{tab1}). While the second phase of LAMOST is ongoing, this work is based on the completed first phase of LAMOST, in particular DR 3, 4, and 5.

\begin{table}
{\tiny
\begin{center}
\caption{Updated spectral statistic of each data release for first phase.\newline\newline
Note: STAR, GALAXY, QSO and UNKNOWN are class assigned by LAMOST pipeline. Numbers may change as spectral reduction and classification pipelines update.}
\label{tab1}
\begin{tabular}{p{0.4cm} p{1.5cm} c c c c c}\hline
Survey  & DATE &STAR & GALAXY & QSO &  UNKNOWN  & TOTAL \\
\hline
Pilot   & 2011-10-24  2012-06-17 & 837,056 & 8,045 &  1,227  & 118,060 & 964,388 \\
DR1     & 2012-09-28  2013-06-03 & 1,536,045 & 12,734 & 6,035 & 127,198  & 1,682,012\\
DR2     & 2013-09-10  2014-06-03 &1,504,329 & 30,432 & 6,382 & 91,399  & 1,632,542\\
DR3     & 2014-09-10  2015-05-30 & 1,516,147 & 26,288 & 8,753 & 88,956 & 1,640,144 \\
DR4     & 2015-09-12  2016-06-02 & 1,551,394 & 39,498 & 13,954 & 96,680 & 1,701,526 \\
DR5     & 2016-09-09  2017-06-16 & 1,226,472 & 36,093 & 14,782 & 119,885 & 1,397,232\\\hline
First phase & 2011-10-24 2017-06-16 & 8,171,443 & 153,090 & 51,133 & 642,178 & 9,017,844\\\hline
\end{tabular}

\end{center}
}
\end{table}

\begin{table}
{\small
\begin{center}
\caption{Classification for the 3522 spectra of 3069 sources.\newline
}
\label{tab_stat}
\begin{tabular}{c c c}\hline
Type  & Spectra & Sources\\
\hline
DA/DA:   & 2625  & 2281 \\
DB/DB:   & 182  & 166 \\
DC/DC:   & 62 & 58 \\
DCA/DCQ     & 2  & 2 \\
DAH/DBH/DAP     &  34 & 31\\
DZ/DZ:    & 36 & 33 \\
DZA/DZB & 6 & 4\\
DAZ/DAZe & 6 & 5\\
DBAZ/DBAZ:  & 3 & 2\\
DBZ/DBZA &  7  & 4\\
DAB/DBA/DBA:   & 76  & 64 \\
DO/DAO/DOA/DBOA$^{a}$    &  23  & 20 \\
DQ/DQ:   &   19  & 19 \\
CV/CV:    &   130  &  106\\
(DA,DB,DC,DAH,DBA)+M/DA+(K,DQ)$^{b}$   &      311 & 274 \\\hline

\end{tabular}
\begin{tablenotes}
\item 
Notes: \newline
Spectral types listed in this paper followed the definitions in Section 2.2 from \cite{Kepler2019}. \newline
Especially, notation ":" means uncertainty mainly due to low S/N.\newline
Notation "e" means emission line present in the spectrum.\newline
$^{a}$DBOA type means its a Helium dominated white dwarf with He II 4\,686 \AA\,line and mild Balmer absorption lines.\newline
$^{b}$White dwarf binary. White dwarf with M, K type star or DQ white dwarf.\newline
\end{tablenotes}
\end{center}
}
\end{table}

\subsection{DA selection}

The raw spectral data are processed with the LAMOST 2D pipeline \citep{Luo2015}. The standard procedures of dark current subtraction, bias subtraction, cosmic ray removal, 1D spectral extraction,  sub-exposures combination are performed in the process of 2D pipeline. Next, the 1D pipeline is used to perform spectral classification, then calculate the radial velocity for each spectrum. Spectral classification adopted by 1D pipeline is done by template matching. The most important part is to construct a spectral classification template library. The early version of this library for LAMOST is built firstly by excluding outliers using local outlier probabilities, then principal component analysis was used to reconstruct spectra. More description can be found in \cite{Wei2014}. The updated version of this library is constructed through clustering algorithm \citep{Kong2019}. The pipeline classification determines object class (STAR, QSO, GALAXY and UNKNOWN), and subclass, e.g.\ (B2, G6, WD). With better pipeline development and more information, there are more correctly classified spectra in DR5. We successfully identify numerous false positives from our previous catalog, mostly spectra with low signal to noise ratio, which we marked ":" for uncertain about the classification (presented in Table\,\ref{tab5}). 

Machine learning has drawn increasing attention in astronomy, where spectral classification is an important potential application \citep{Jiang2013,Liu2014,Li2018}. Thus we applied machine learning algorithm to select DA candidates in this work, as another independent method besides adopting pipeline subclass. There are various types of machine learning algorithms, e.g.\ support vector machine, Bayesian networks, decision tree learning. Based on our previous experience performing spectral classification \citep{Bai2018,Bai2019}, we choose an algorithm called random forest \citep[RF]{Breiman2001}. It performs better than other algorithms, in terms of time cost and accuracy. 

The RF algorithm operates by constructing a multitude of decision trees at training and outputting the class that is the mode of the classification of the individual trees \citep{Breiman2001}. Random decision forests correct for decision trees's habit of overfitting to their training set. The RF algorithm implementation used in this work is a supervised algorithm, and adopts 100 trees (estimators). All flux values of a spectrum are used as parameters. The largest weight is given to parameters in the Balmer line wings. For each spectrum, a probability value will be produced by the RF algorithm. Then a simple binary classification is used. If the probability of being a DA star is greater than 50\%, this object will be classified as a DA candidate. Otherwise, it will be classified as non-DA. This algorithm implementation is from scikit-learn python package \citep{Pedregosa2011}, and information of the specific model can be found in their website \footnote{\url{https://scikit-learn.org/stable/modules/generated/sklearn.ensemble.RandomForestClassifier.html}}. Considering the fact that observed spectra, even for the same source, vary from different surveys (i.e. for the same faint source, SDSS spectra are likely to have higher signal-to-noise (S/N) than LAMOST spectra. And SDSS spectra are better flux calibrated, compared to LAMOST spectra etc.), we first built a DA training sample from LAMOST spectra. First of all, there are 379 DA spectra of S/N greater than 10 in SDSS $g$ band selected from our previous catalogue \citep{Guo2015b}. Secondly, we randomly selected 100 spectra with the same S/N limit for each spectral type classified by LAMOST pipeline, which is around 6k of non-DA spectra in total. Together, these 379 DA spectra and $\sim$ 6k non-DA spectra are used to train the model, then the model is applied to a randomly selected 50k spectra sample for initial test. Based on our test result, most of the contaminants come from A stars. It is understandable, because our RF algorithm mostly uses the broad and deep Balmer line profile of DA stars to select candidates. Some A stars have similar profiles in Balmer line regions. Thus, we added an additional training model specific for separating DA and A stars. The main idea is to distinguish A stars by using their multiple absorption line features in the near-infrared region. Therefore, the largest weight is given to that region. This time, same 379 known DA stars and another 1\,000 A type star spectra (classified by LAMOST pipeline) with S/N above 10 are selected to train the model. Then this additional model is applied to DA candidates identified by first model. It produces a probability value as well. But this time, if the probability is greater than 50\%, this object is classified as A type star. Consequently, its label is changed from DA to non-DA. At last, this code is applied to the whole data set in LAMOST DR 3, 4, \& 5. In total, there are 6\,662 unique spectra candidates selected by our RF algorithm. Each spectrum is visually inspected by eye, 1\,710 DAs are identified.

 In addition, we used a conventional colour-colour cut (Formula 1-4 in \citealt{Eisenstein2006}, and Table 1 in \citealt{Girven2011}) as supplementary methods to select DAs. Even though LAMOST is basically a spectroscopic survey, there are imaging surveys like XSTPS-GAC and SDSS, providing $g$, $r$, $i$ magnitudes for more than 50\% of the input catalogue entries. Around 1.46 million entries in LAMOST DR 3, 4 \& 5 have $u$, $g$, $r$, $i$, and $z$ magnitudes, mostly from SDSS. Using method from \cite{Eisenstein2006}, 43\,461 white dwarf candidate spectra are selected, while 847 DA star candidate spectra are selected adopting method from \cite{Girven2011}. For pipeline selection, we select spectra that are classified by the pipeline as WD, WDMagnetic, DoubleStar, or CarbonWD as DA candidates. Spectral types of WDMagnetic, DoubleStar, and CarbonWD are also included, because small number of DA can be misclassified by the pipeline and identification of those types of WD is a part of our goals as well. Those methods result in 6\,662 unique candidate spectra from RF algorithm, 43\,461 candidate spectra from \cite{Eisenstein2006}, 847 DA candidate spectra from \cite{Girven2011}, and 7\,024 DA candidates from pipeline selection. In total 2\,620 bona fide DA spectra and many various types of WDs are identified after visual inspection (See Table. \ref{tab_stat}). After cross-matching with the literature, 393 DAs are new identifications \citep{Kleinman2004,Kleinman2013,Zhao2013,Zhang2013,Guo2015b,Kepler2015,Kepler2016,Kepler2019}.

One should note that our strategy in selecting DA stars is to include as many DAs as possible with manageable candidate size. This may result in considerable contamination rate, but rely on visual inspection, contaminants are able to be removed and a relatively complete DA sample can be obtained. Thus, we emphasis the importance of visual inspection of candidate spectra in this work. According to our identification, the efficiency (defined as the ratio of correctly identified DAs to the total number of objects identified as DA candidates) of LAMOST pipeline is $2\,234/7\,024=31.8\%$. But more accurately, there are 3869 spectra classified by LAMOST pipeline as DA (i.e. subcalss WD). Spectra classified as subclass WDMagnetic and DoubleStar are also included in pipeline method to ensure Magnetic WDs and WDMS binaries are included in our main catalogue. Therefore, the actual efficiency of LAMOST pipeline should be $2\,234/3\,869=57.7\%$. The main contaminants are low S/N spectra. Of the 3\,869 spectra classified as subclass WD, only 928 spectra have S/N greater than 5 in $u$, $g$, $r$, $i$, and $z$ bands. Substantial type A and F stars with very noisy spectra have been wrongly matched with DA templates. The efficiency of RF algorithm is $1\,710/6\,662=25.7\%$. Similarly, the main contaminants are still A stars (3\,645 of 6\,662), together with some F stars and hot subdwarfs. Even though the efficiency of Eisenstein colour cut is only $1\,172/43\,461=2.7\%$, majority of other WD types (except DA and DB) are identified through this method. Contaminants are various, more than half of the candidates selected by Eisenstein colour cut method are QSOs, along with B, A, F stars and other types of WDs. The efficiency of Girven colour cuts is $373/847=43.9\%$. The main contaminants are type G, A, B stars and QSOs. There are 216 DAs missed by LAMOST pipeline and RF algorithm. After a closer look at those DAs, about 100 DAs are found to be misclassified by pipeline as A stars mostly, and B, F stars. Those DAs display relatively narrow absorption lines, compared to the majority of DAs. But they are indeed located in the WD region in the colour-colour plot and $Gaia$ colour-magnitude diagram. This indicates the DA templates used by LAMOST pipeline lack low mass DAs. Same problem exists in the RF method, since its training sample is based on our previous catalogue, and identification of low mass DA is generally difficult before $Gaia$. Apart from those 100 missed DAs, another 100 missed DAs have very low S/N spectra (S/N<5 in all SDSS bands). It appears that broad absorption line features in those very noisy spectra have not been picked up by template matching and RF algorithm.

Regarding the completeness ratio of our identified DA sample to all DAs observed by LAMOST, we applied simple and rough ways to estimate, aiming to evaluate the robustness of our DA identification methods. From four methods, LAMOST pipeline identified 2\,234 DA spectra, two colour cut methods identified 753 spectra, and RF algorithm identified 1\,710 spectra. Most DA spectra are discovered by two or more methods. Because there is no independent LAMOST imaging survey, substantial LAMOST sources have no required colour to perform the selection, DA spectra identified via colour cut method is highly incomplete. Therefore, only pipeline and RF algorithm identified DAs are used to estimate the completeness. There are 1\,819, 1\,523, and 2\,106 DA spectra with S/N no less than 5 in $g$ band identified by LAMOST pipeline, RF method and four methods combined, respectively. The number of common spectra identified by both pipeline and RF is 1\,379. Thus, the percentage of pipeline missed DA spectra is about $(2\,106-1\,819)/1\,819=18\%$. Since pipeline and RF together identified 1\,963 unique DA spectra, the percentage of pipeline and RF missed spectra is about $(2\,106-1\,963)/1\,963=7\%$. Therefore, the total missed DA spectra could be $2\,106\times18\%+2\,106\times7\%=527$, which means a lower limit of completeness should be around $2\,106/(2\,106+527)=80\%$. With the help of pipeline and RF method, we managed to identify 1\,963 DA spectra with S/N no less than 5 in $g$ band, which means an upper limit of completeness of pipeline and RF combine is 1\,963 /\,2\,106=93\%. In summary, the completeness ratio of identified DA sample to all DAs observed by LAMOST is in the range of 80\% to 93\%. It shows the estimated completeness is consistent, both indicate the number of DA spectra that are not included in our catalogue is not significant.

Another separate test has been carried out by cross matching LAMOST DR 3, 4, \& 5 spectra with most recently published SDSS DR 14 DA catalogue first \citep{Kepler2019}, there are 272 DA common sources with S/N ratio no less than 5 in $u$, $g$, and $r$ band. Then those sources were cross matched with our DA catalogue, five spectra were found to be missing in our DA catalogue. Therefore, the completeness estimated this way is $5/272=98\%$, consistent with our previous estimation. Note that those five missed sources were added to our catalogue, making the total number of identified DA spectra 2\,625.

To further illustrate the ability of different methods in identifying DA spectra as a function of S/N, Fig. \ref{fig01} is here to show the statistics of DA spectra identified by LAMOST pipeline and RF versus S/N of LAMOST spectra in SDSS $g$ band. DA spectra identified by colour cut methods are not considered here as their sample are biased by lacking photometric data. All panels have shown that pipeline method are better than our RF method in identifying more DA spectra and yielding less contaminants, regardless of S/N. Bottom panel indicates that pipeline method yields much more contaminants at low S/N than at high S/N, while the contamination rate of RF method seems not much affected by S/N of LAMOST spectra. Nonetheless, more than 60\% of RF method selected candidates are contaminants.

\begin{figure}
\center
\includegraphics[angle=0,width=0.5\textwidth]{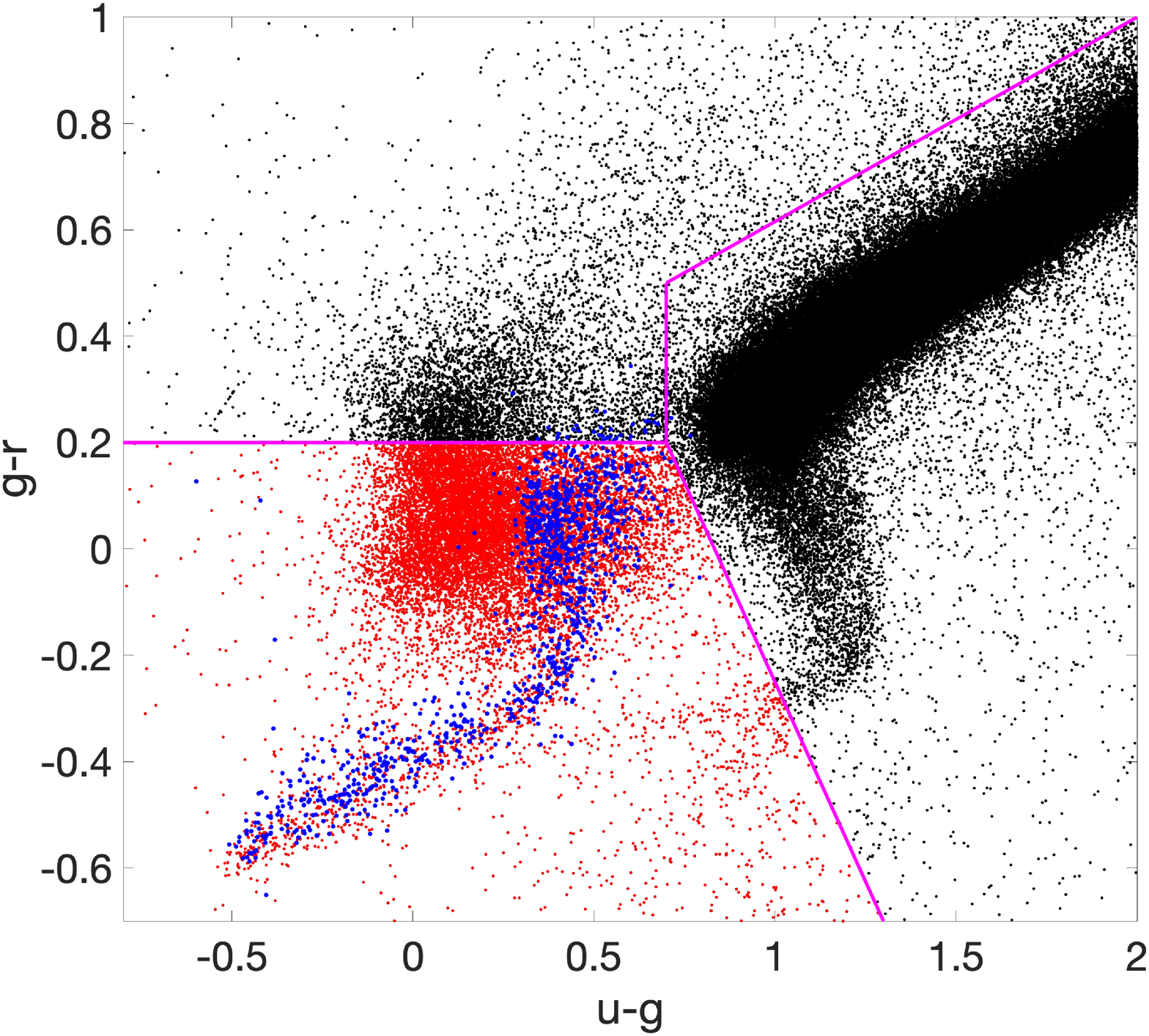}
\caption{The colour-colour cut selection, based on SDSS $u$-$g$ and $g$-$r$, de-reddened. The black dots are LAMOST sources with $u$, $g$, $r$ magnitudes, while red dots are WD candidates selected following \protect\cite{Eisenstein2006}, blue dots are DA stars selected following \protect\cite{Girven2011}. Magenta lines have separated different regions defined by \protect\cite{Eisenstein2006}. Bottom left is WD candidate region, while right region is main-sequence and blue horizontal branch region. Black dots above the horizontal magenta line are mostly quasars.}
\label{fig00}
\end{figure}

\begin{figure*}
\center
\includegraphics[angle=0,width=1.0\textwidth]{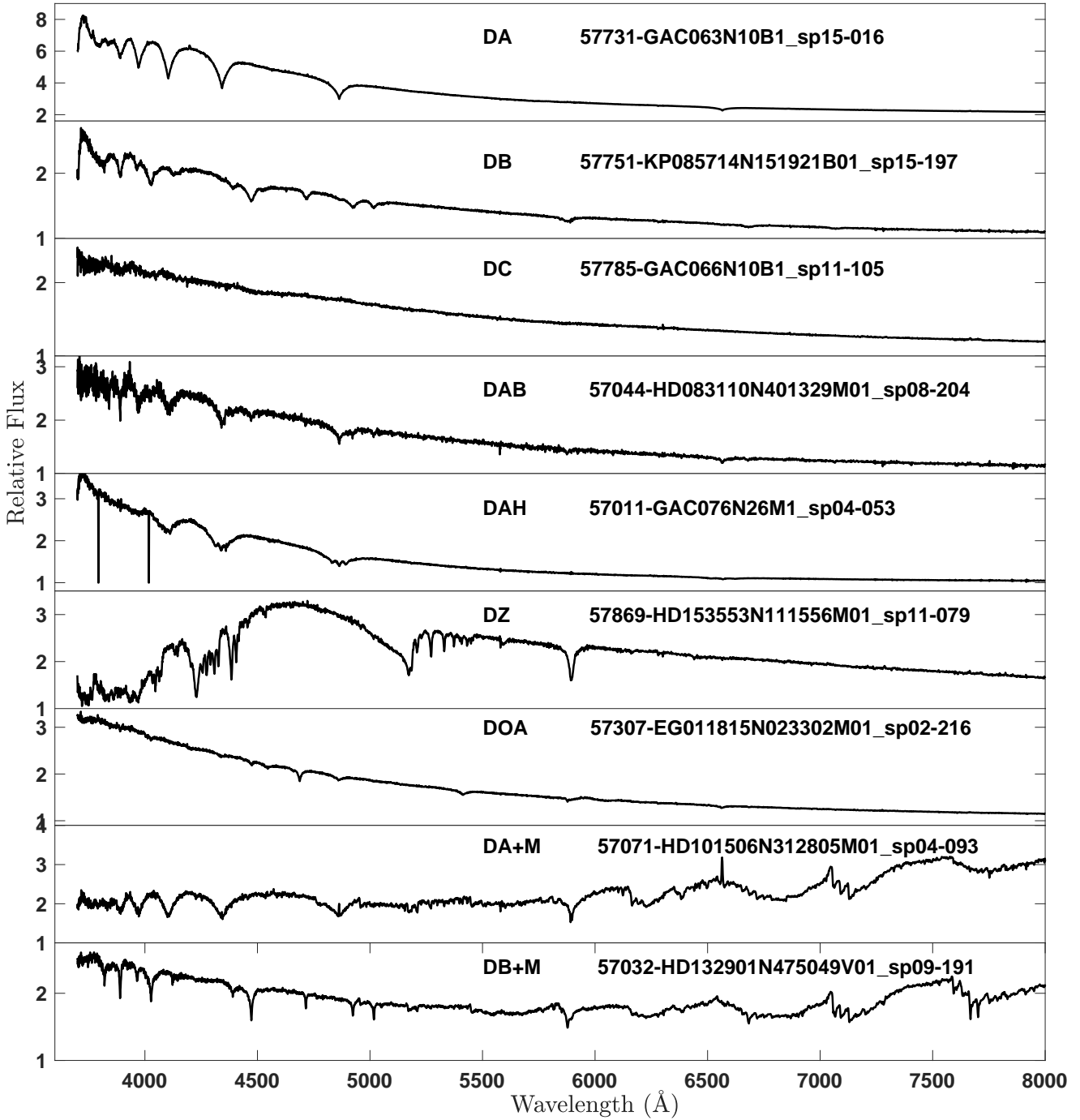}
\caption{Examples of various WD spectral types identified in LAMOST. In order to focus on the relevant spectral features, this figure only shows the wavelength range 3500--8000\,\AA. Classified types and their unique LAMOST spectral IDs are also shown.}
\label{fig0}
\end{figure*}

\begin{figure}
\center
\includegraphics[angle=0,width=0.5\textwidth]{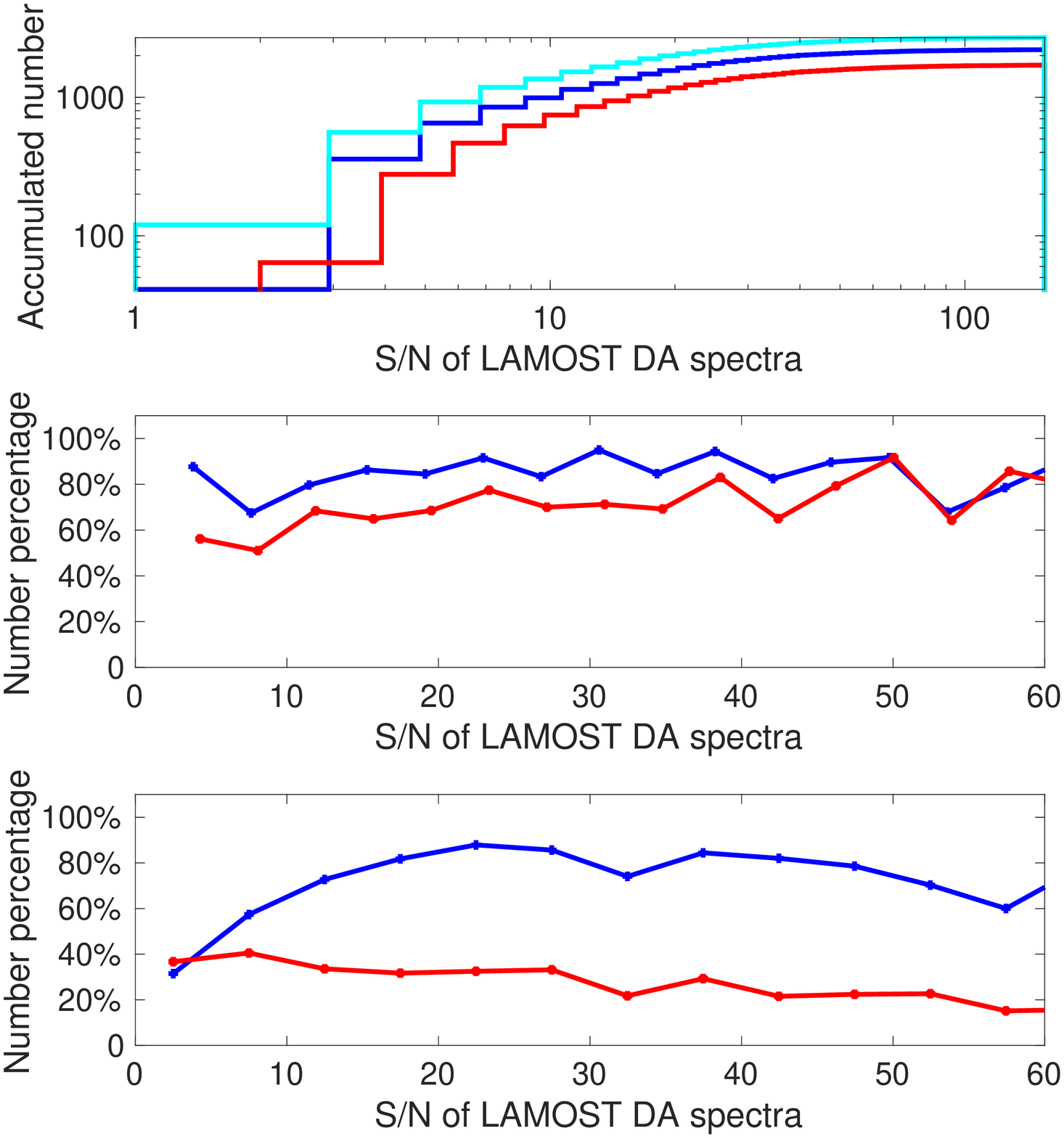}
\caption{The statistics of DA spectra number identified by different methods as a function of S/N in SDSS $g$ band. Top panel: Accumulated number histograms of LAMOST pipeline, RF algorithm, and all methods identified DA spectra in blue, red and cyan colour, respectively. Middle panel: Number percentage of LAMOST pipeline (blue line) and RF algorithm (red line) identified DA spectra in all DA spectra identified as a function of S/N. Bottom panel: Blue line represents number percentage of LAMOST pipeline identified DA spectra in candidate spectra selected by pipeline, while red line represents number percentage of RF algorithm identified DA spectra in candidate spectra selected by RF method. }
\label{fig01}
\end{figure}

\subsection{DB selection}

Because there are no DB templates in the model database of LAMOST classification pipeline, we only applied RF machine learning method to identify DB stars.

There are 1\,842 objects in SIMBAD that are listed as DB stars. Those known DBs in SIMBAD were cross-matched with all spectra from LAMOST DR5, resulting in 178 matches, where 150 of these are classified as UNKNOWN by LAMOST pipeline. A large portion of those 150 spectra have very low S/N ratio that are not able to be identified by eyes. Thus, 58 of 178 known DBs are not recovered by our DB selection. Because we can not identify them based on their unrecognisable noisy LAMOST spectra alone, and these noisy spectra have no use for any analysis, those are not included as DBs in this paper. In order to select a high quality sample to best represent DB features, we only chose those LAMOST spectra with S/N $>10$ in the SDSS $g$ band. There are 45 spectra that meet this criterion and form the training sample. Next, after removal of the 178 known SIMBAD DB spectra, roughly 1\,000 spectra with S/N>10 were randomly selected from LAMOST DR5. Most of those 1\,000 spectra are different types of star, together with galaxies and QSOs. DB spectra are ensured to be excluded by careful spectral inspections. Similar to RF method applied to identify DAs, we used all unsmoothed 45 DB and $\sim$1\,000 non-DB spectra to construct the supervised DB classifier for the RF algorithm. All flux values of a spectrum are adopted as parameters. The largest weight is given to parameters in the region of Helium absorption line wings. A simple DB or non-DB classification is used. An object is classified as a DB candidate, when its probability of being DB star is greater than 50\%. Or this object will be classified as non-DB. In the final step, all spectra in DR\,3, 4 \& 5 are fed to the classifier, DB candidates are obtained.

There are 12\,572 DB candidates selected by the RF algorithm. After visual inspection, 182 spectra of 166 sources were found to be bona fide DBs. After cross-matching with previous published catalogs, 46 DBs are new identifications \citep{Kleinman2004,Kleinman2013,Kepler2015,Kepler2016,Guo2015b}. We believe that higher contamination rate in DB selection is caused by smaller training sample size relative to DA training sample. In addition, a second training is applied to distinguish DA and A star in DA machine learning selection.

\begin{figure}
\center
\includegraphics[angle=0,width=0.5\textwidth]{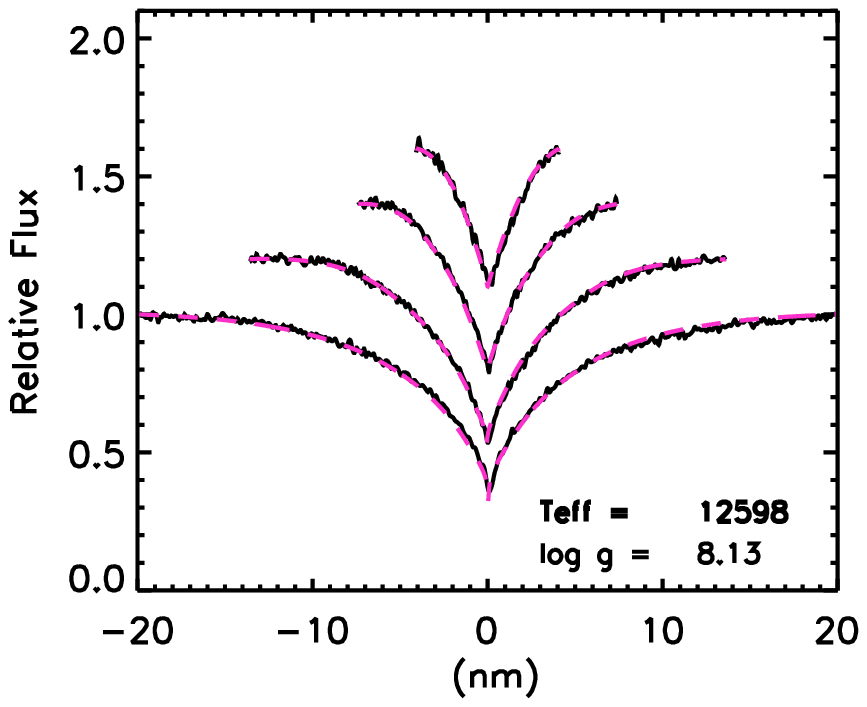}
\includegraphics[angle=0,width=0.5\textwidth]{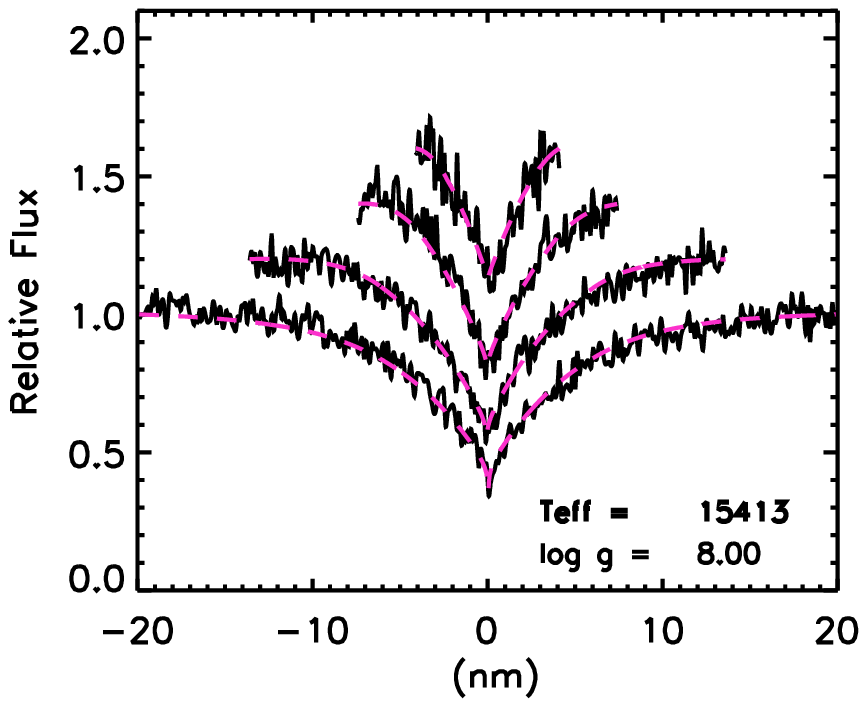}
\caption{Balmer line fitting examples for J041010.37+180222.8 (upper panel, S/N=107) and J071004.93+292403.2 (lower panel, S/N=28). For both cases, the black solid lines are observed and normalised DA spectra, while red dashed lines are the best fit model. From bottom to top of each panel, fitted absorption lines are H$\beta$, H$\gamma$, H$\delta$, and H$\epsilon$, respectively.}
\label{fig1}
\end{figure}

\subsection{Other WD spectral types}
Besides DA and DB stars, other spectral types of WD are identified as well, as a result of our effort in searching for DA stars. They are from the colour-colour cut selection method from \cite{Eisenstein2006}, \cite{Girven2011} and visual inspection of spectra that are classified as WDMagnetic, DoubleStar and CarbonWD. Colour-colour cut from \cite{Eisenstein2006} yielded more types of WD than any other methods. But not all LAMOST sources have SDSS photometry, so WD types except DA and DB are highly incomplete.

There are 5 DAZs, 4 DBZs and 2 DBAZs identified in this work. Two of five DAZs are classified as WD by LAMOST pipeline, Three of four DBZs and one of two DBAZs are selected as DB candidates by RF algorithm. The rest $3+1+1=5$ are discovered by Eisenstein's colour cut method. Based on a rough estimation, the fraction of those special metal polluted WDs in Eisenstein's colour cuts selected WD candidates is about $5/43\,461=0.011\%$, and the fraction of this colour cuts selected candidates in LAMOST DR3, 4 \& 5 that have SDSS photometry is $43\,461/1\,455\,566=2.98\%$. Therefore, the number of those special WDs that have been missed in LAMOST because they do not have SDSS photometry is $(4\,740\,458-1\,455\,566)\times(0.011\%\times2.98\%)\approx11$. Besides those three types of WDs, magnetic WDs (DAH, DBH, and DAP) mainly come from pipeline classified WDMagnetic, while pipeline classified DoubleStar identified most WDMS binaries and CVs. And all the spectral types identified in this work with their corresponding numbers are listed in Table.\ \ref{tab_stat}. A small part of basic parameters of all WDs identified in LAMOST DR 3, 4 \& 5, including designation, ra, dec, observation date, unique spectral ID, and WD subtype, has shown in Table.\ \ref{tab2}.

\section{Parameter determination}

\subsection{$T_{\rm eff}$ and log $g$ for DA stars}

In order to ensure the accuracy of the derived parameters for DAs, only spectra with S/N $>15$ in SDSS $g$ band are fitted. For sources with multiple observations, those with highest $g$ band S/N are used. To estimate the effective temperature and surface gravity of DA stars, absorption line profiles from H$\beta$ to H$\epsilon$ were fitted to theoretical spectral models. Before performing the spectral fitting, both the observed and model absorption line profiles were trimmed and normalised following standard procedures \citep{Liebert2005}. The next step is to use a minimisation technique to fit the line profile. The technique adopted here is the well known non-linear least-squares method of Levenberg-Marquardt, that is based on a steepest descent method. To fit the best model template, the open source {\sc idl} package {\sc mpfit} was used \citep{Markwardt2009}. The DA atmosphere models used here are provided by D.\ Koester (2015, private communication) and are based on those described in \cite{Koester2010}. The errors for $T_{\rm eff}$ and log $g$ were determined by stepping the parameters about the minimum $\upchi^{2}$. The difference, which is calculated between each current minimum $\upchi^{2}$ and the previous true minimum $\upchi^{2}$, corresponds to 1$\sigma$ for a given number of free model parameters is regarded as the error.

\begin{figure}
\center
\includegraphics[angle=0,width=0.53\textwidth]{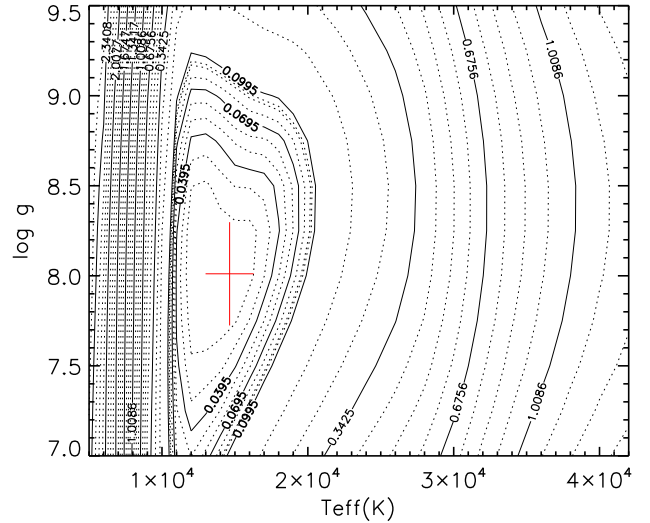}
\caption{$\upchi^{2}$ contour plot of $T_{\rm eff}$ and log\,$g$ determination for DA J041010.37+180222.8. Red cross represents final result with its uncertainties.}
\label{fig2}
\end{figure}

Notably, we introduced {\em Gaia} $G\rm_{BP}$-$G\rm_{RP}$ colour into the spectral fitting process. First, a sample of SDSS identified DAs with relatively high precision $T_{\rm eff}$ is constructed. Next, an empirical relation is established between $T_{\rm eff}$ and their corresponding {\em Gaia} colour. Then for a new DA spectrum to be fitted, an initial effective temperature value was given by the empirical relation, based on its {\em Gaia} colour. The initial value was then used as one input parameter in {\sc mpfit} code to start model fitting iterations (More details in Zhang et al. in prep). After this preparation, it's the same pure conventional spectral fitting to theoretical model spectra to obtain the final $T_{\rm eff}$ and log $g$. By using {\em Gaia} data in this way, the final effective temperature will be determined more robustly and accurately than our previous study \citep{Guo2015b}. Two examples of Balmer line fitting are shown in Fig.\ref{fig1}. For comparison purposes, we chose one spectrum with high S/N (J041010.37+180222.8), and another spectrum with low S/N (J071004.93+292403.2). Figure \ref{fig2} shows a $\upchi^{2}$ contour plot of $T_{\rm eff}$ and log $g$ for the source J041010.37+180222.8. The best-fitted values for $T_{\rm eff}$ and log $g$, together with their uncertainties are estimated from the minimum converged residual $\upchi^{2}$. The median error of $T_{\rm eff}$ and log $g$ in our DA sample is roughly 8\% and 4\%, respectively.

\subsection{$T_{\rm eff}$ and log $g$ for DB stars}

\begin{figure*}
\center
\includegraphics[angle=0,width=0.7\textwidth]{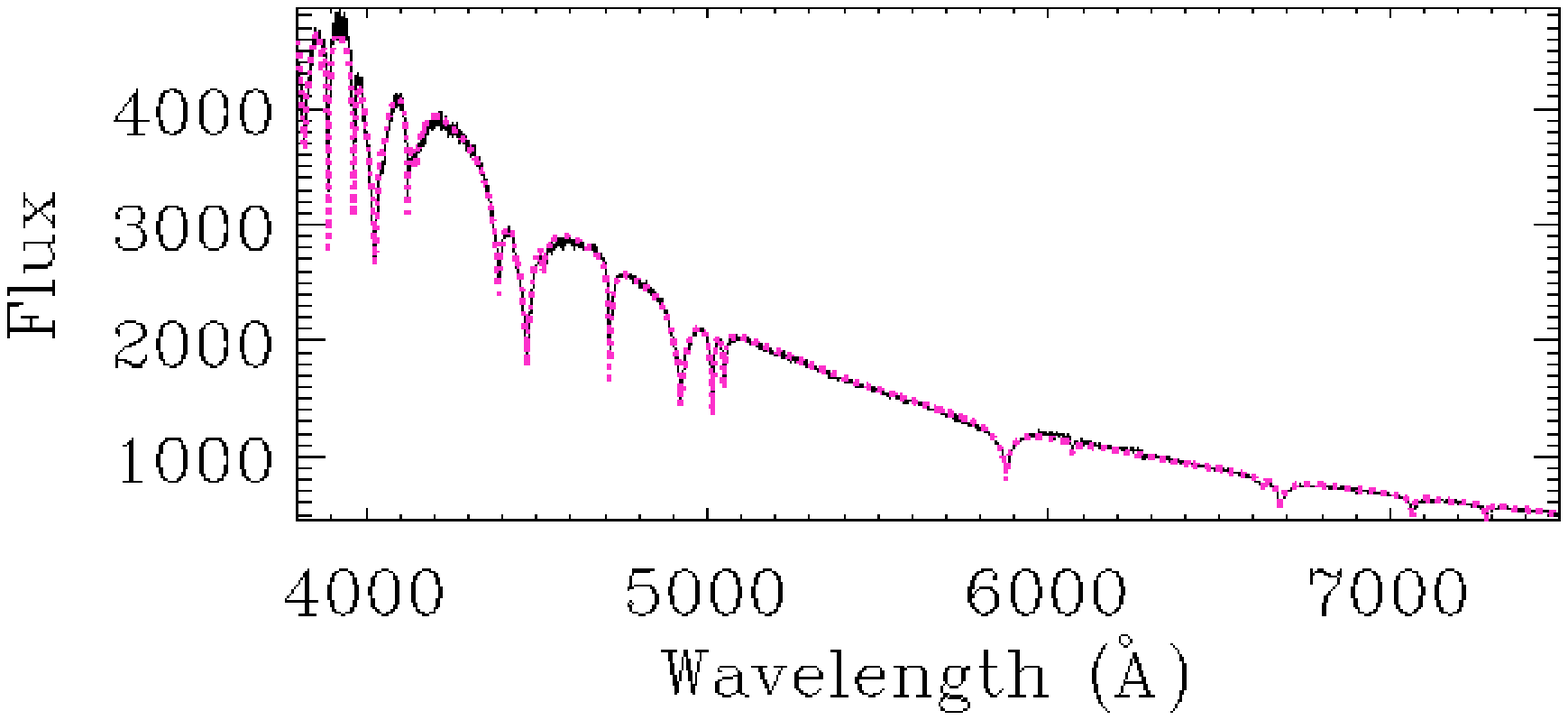}
\includegraphics[angle=0,width=0.7\textwidth]{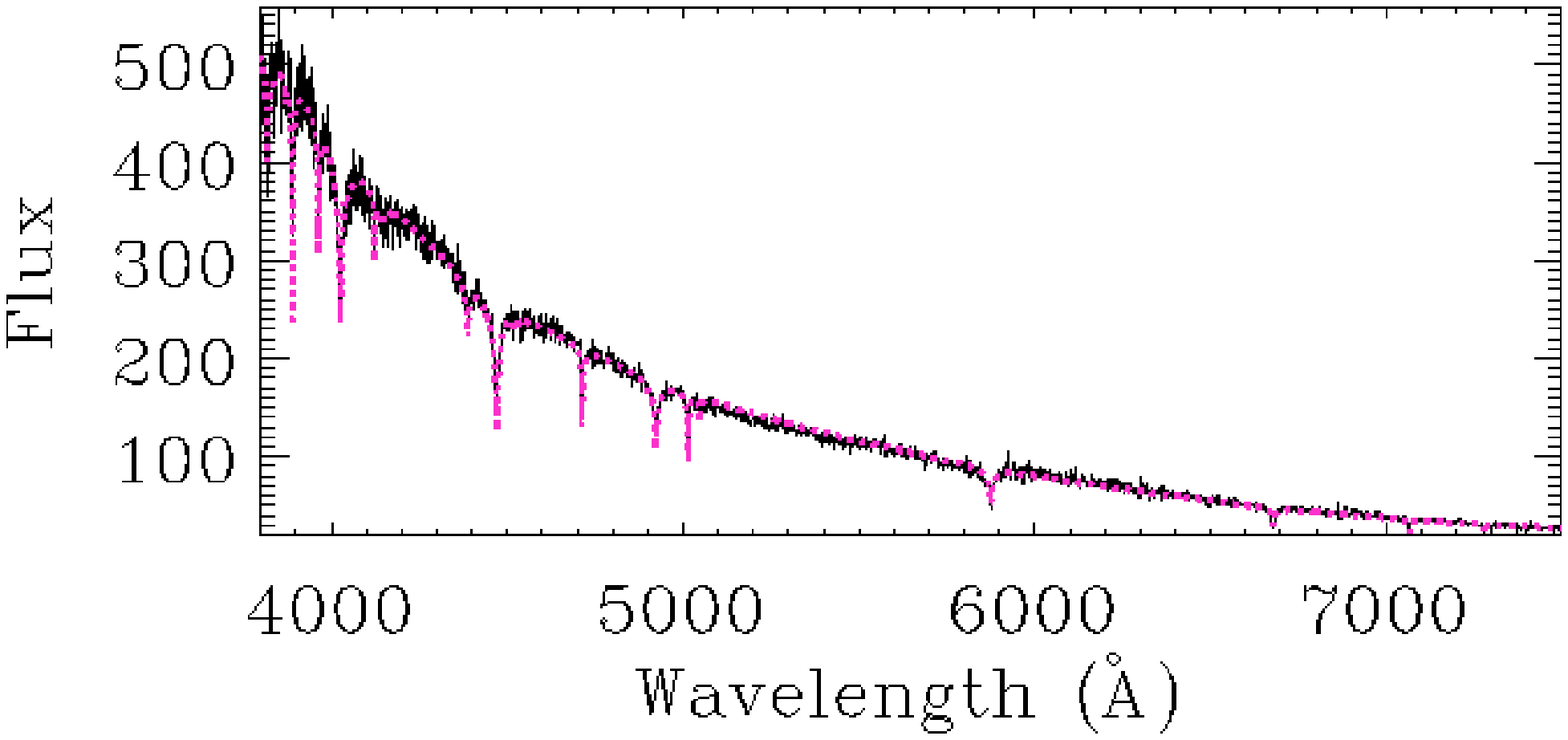}
\caption{DB Helium absorption lines model fit comparison between J164718.38+322832.8 (upper panel, S/N=57) and J012148.23-001053.0 (lower panel, S/N=29). The black solid lines are observed DB spectra, while red dashed lines are best fit model spectra.}
\label{fig3}
\end{figure*}

To ensure the number of DB stars with parameters that can be compared with the literature, we adopt a S/N limit of 10. The DB spectrum with highest S/N was chosen for model fitting in the case that a source has multiple spectra. Also, because there can be a large number of helium absorption lines, the fitting process is complicated by their normalisation. Therefore, the parameter determination for DB spectrum is usually done by fitting the entire spectrum to the theoretical model. But in this case, flux calibration errors may affect the accuracy of parameter estimation. Flux calibration is difficult for LAMOST spectra, owing to e.g.\ flat-fielding over large fields of view, use of pseudo-standard stars, and possible unknown reddening \citep{Du2016}. To evaluate which is more suitable for LAMOST DB spectral fitting, by fitting only Helium absorption lines after continuum normalisation (similar to DA spectral fitting) or fitting the entire flux calibrated spectrum directly, a sample of a few tens of known DB stars from \cite{Koester2015} is selected to perform a less formal test. Those DB stars are common sources of our LAMOST DB sample and DB catalogue from \cite{Koester2015} with relatively small errors of derived $T_{\rm eff}$ and log $g$. Two sets of parameters are derived from these two different methods, who performed on LAMOST spectra. Next, both results were compared with SDSS catalogue values, and we found that fitting the entire flux calibrated spectrum yielded more consistent results. Therefore, we chose this method despite flux calibration concerns. The DB models used in the fits were pure He models, kindly provided by \cite{Koester2010}. Fig. \ref{fig3} shows two examples of DB stars where the entire spectra were fitted by models. One is DB spectrum with high S/N (J164718.38+322832.8, S/N=57), while the other has lower S/N (J012148.23-001053.0, S/N=29). The best-fitted $T_{\rm eff}$ and log $g$, as well as their uncertainties are also obtained from the minimum converged residual $\upchi^{2}$. It is noted that the effective temperature and surface gravity for DB training sample  we adopted ranges from 11\,722\,K to 33\,894\,K and 7.6 to 8.7 dex, respectively, according to SIMBAD records. Therefore, even though these ranges should cover most of DB stars, it is still possible that a small number of DBs with $T_{\rm eff}$ and log $g$ outside these ranges are missing in our DB catalogue.

\subsection{Mass and cooling age}

Once the $T_{\rm eff}$ and log $g$ are obtained from the model fitting, their mass and cooling age can be estimated based on evolutionary models \citep{Fontaine2001} \footnote{The cooling model can be downloaded from the Web site: \url{http://www.astro.umountreal.ca/\textasciitilde bergeron/CoolingModels/}.}. For DAs, the cooling models in \cite{Wood1995} was used for the carbon-core with thick ($q_{\rm H}= M_{\rm H}/M_{\star}=10^{-4}$) hydrogen layers with $T_{\rm eff}$ greater than 30,000 K. For models with effective temperature less than 30,000 K, the cooling models for carbon-oxygen cores in \cite{Fontaine2001} with thick ($q_{\rm H}=10^{-4}$) hydrogen layers are used. For DBs, the cooling models in \cite{Wood1990} was adopted for the carbon-core with thick ($q_{\rm He}= M_{\rm He}/M_{\star}=10^{-4}$) helium layers \citep{Bergeron2001}.

\begin{figure}
\center
\includegraphics[angle=0,width=0.5\textwidth]{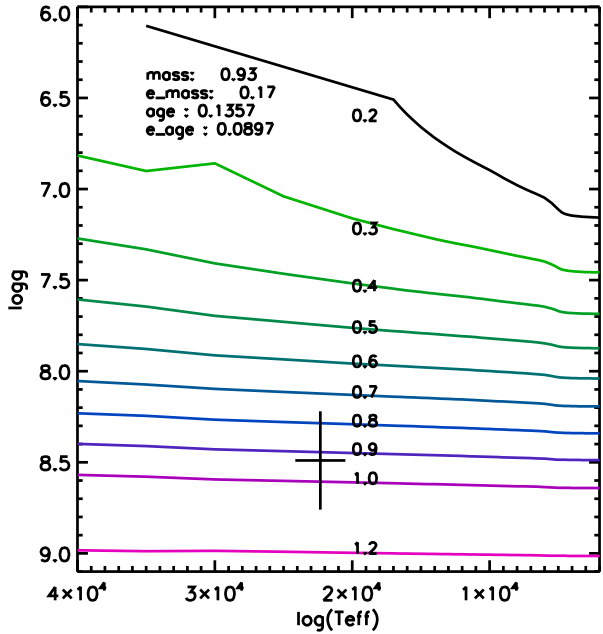}
\caption{One example of mass and cooling age determination, based on effective temperature and surface gravity. Black cross represents the location of this example with uncertainties in the effective temperature vs. surface gravity space. Different lines represent theoretical models with correspondent mass. Determined mass and cooling age are shown in the upper left with their errors.}
\label{fig34}
\end{figure}

\begin{figure}
\center
\includegraphics[angle=0,width=0.5\textwidth]{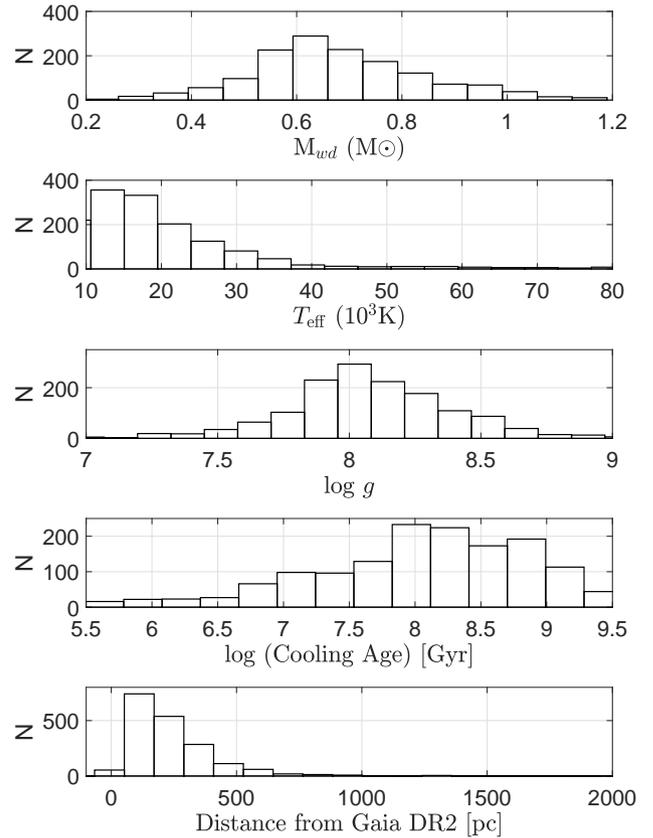}
\caption{Distributions of derived DA parameters, including DAs with updated parameters from LAMOST DR2. From top to bottom: mass, effective temperature, surface gravity, cooling age and inverted parallax from Gaia DR2.}
\label{fig4}
\end{figure}

An example of mass and cooling age determination is demonstrated in Figure \ref{fig34}. The resulting distributions of mass and cooling age are shown in Figure \ref{fig4} and Figure \ref{fig6}, respectively. There are 1316 DAs and 70 DBs with good quality spectra that can be used to construct these distributions. 

Looking at Fig. \ref{fig4}, the DA mass range is from 0.3\,M$_{\odot}$ to 1.2\,M$_{\odot}$, with a peak near 0.62\,M$_{\odot}$. This is in agreement with the mean mass of 0.613\,M$_{\odot}$ from \cite{Tremblay2011} and our previous research \citep{Guo2015b}. But our peak mass is slightly less massive than the mean mass of 0.649\,M$_{\odot}$ from \cite{Kepler2015}. As is well established in the literature \citep[and references therein]{Rebassa2015} as well as in \cite{Guo2015b}, two obvious, non-single Gaussian distribution features present near 0.4\,M${_{\odot}}$ and 1.0\,M$_{\odot}$. However, these features are not evident in our combined DA mass distribution of DR 1-5.

\begin{figure}
\center
\includegraphics[angle=0,width=0.55\textwidth]{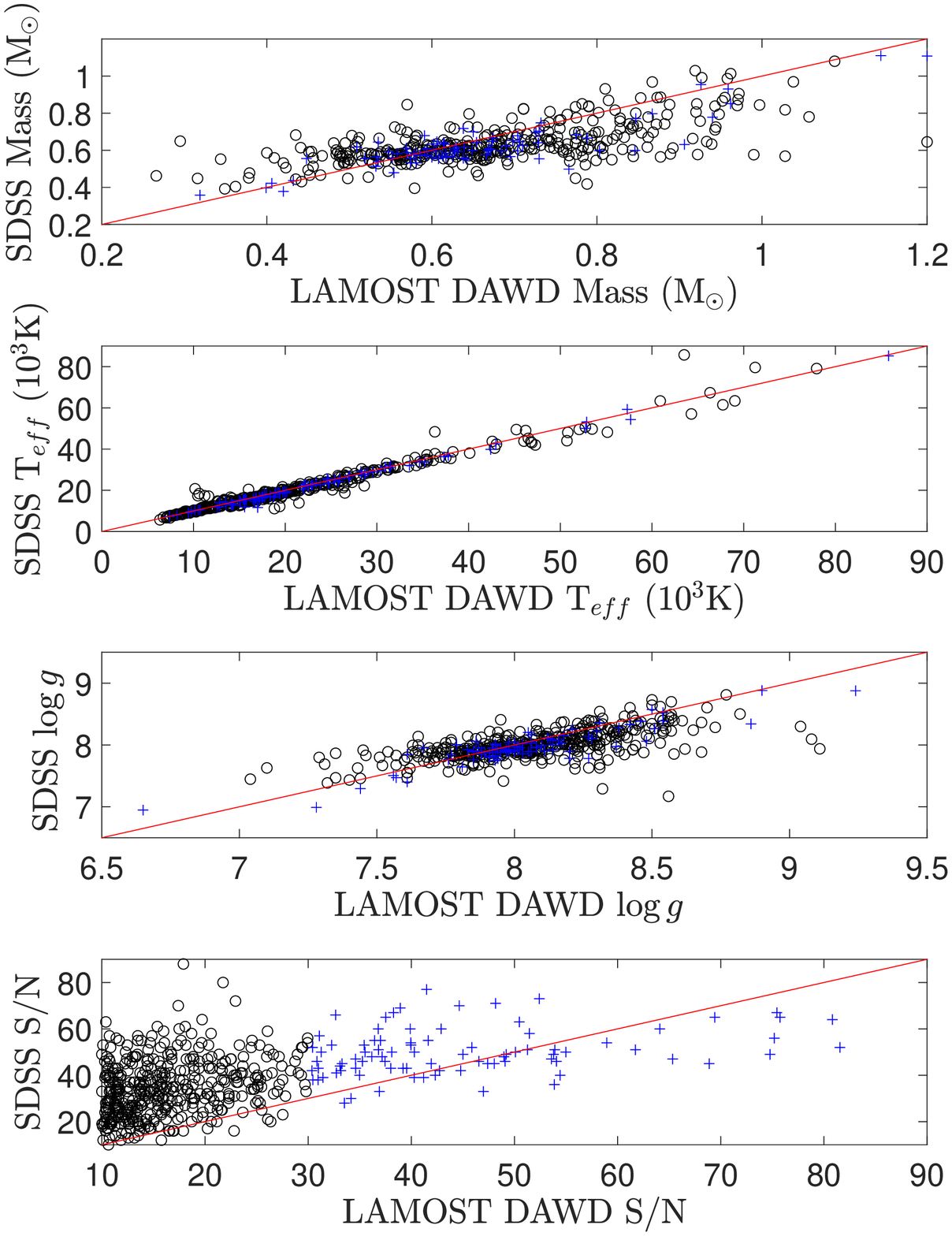}
\caption{Comparisons between LAMOST DA parameters (DAs with updated parameters from LAMOST DR2 are also included) and SDSS parameters from \protect\cite{Kepler2019}. From top to bottom: mass, effective temperature, surface gravity and S/N. Black circles represent DAs with S/N of LAMOST spectra between 10 and 30, while blue crosses are DAs with S/N of LAMOST spectra greater than 30. The red solid lines are unit slope relation.}
\label{fig5}
\end{figure}

\begin{figure}
\center
\includegraphics[angle=0,width=0.55\textwidth]{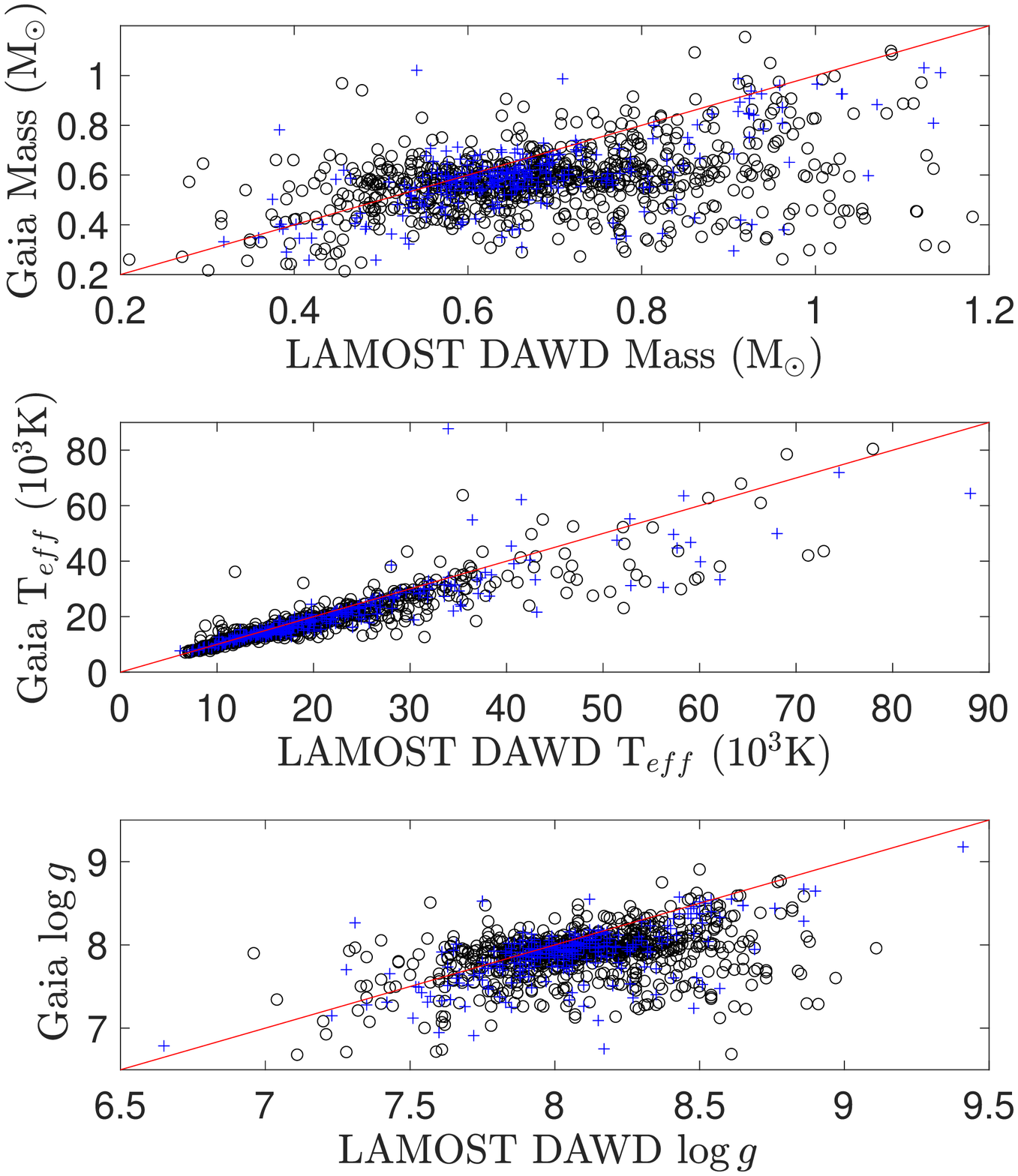}
\caption{Comparisons between LAMOST DA parameters (DAs with updated parameters from LAMOST DR2 are also included) and {\em Gaia} DR2 WD parameters from \protect\cite{Gentile2019}. From top to bottom: mass, effective temperature and surface gravity. Black circles represent DAs with S/N of LAMOST spectra between 10 and 30, while blue crosses are DAs with S/N of LAMOST spectra greater than 30. The red solid lines are unit slope relation.}
\label{fig51}
\end{figure}

\begin{figure}
\center
\includegraphics[angle=0,width=0.5\textwidth]{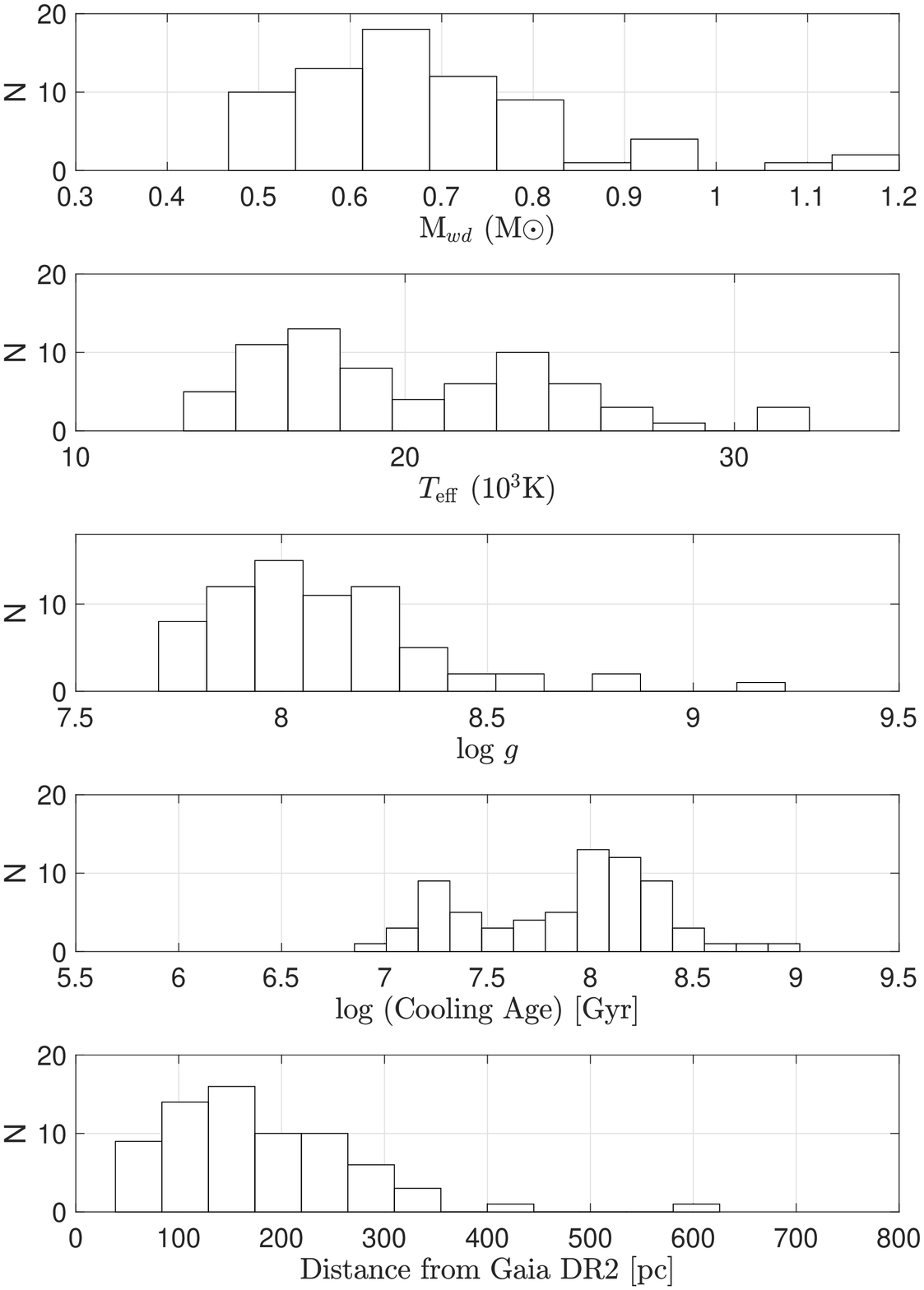}
\caption{Distributions of derived DB parameters. From top to bottom: mass, effective temperature, surface gravity, cooling age and inverted parallax from {\em Gaia} DR2.}
\label{fig6}
\end{figure}

Of all the fitted LAMOST DAs, 506 stars are found to have fitted parameters in the SDSS DR14 white dwarf catalogue \citep{Kepler2019}. In Fig. \ref{fig5}, we compare our results with \cite{Kepler2019}. In general, both sets of results are in good agreement. Sources common to both studies with higher S/N LAMOST spectra exhibit higher consistency between the studies. Also, the derived surface gravities and thus corresponding masses are more discrepant at the high and low ends. Our study systematically underestimates masses at the high end, and overestimates mass at the low end, relative to the SDSS WD study. In terms of S/N, the difference is displayed clearly in the bottom panel of Fig. \ref{fig5}. Regarding DAs with LAMOST spectra S/N of less than $\sim$50, the SDSS spectra quality are obviously better than LAMOST spectra. However, as for DAs with LAMOST spectra  S/N of greater than 50, the LAMOST spectra quality are clearly better. We suggest our derived parameters are reliable for S/N above 30. Similarly, there are 1059 stars found to have fitted parameters in the {\em Gaia} DR2 WD catalogue \citep{Gentile2019}. In Fig. \ref{fig51}, our parameters and results from \cite{Gentile2019} are compared. Results from both catalogues are generally agree with each other. However, the derived surface gravity and mass data are clearly more disperse than the data from comparisons between LAMOST and SDSS, despite S/N of LAMOST spectra. It is understandable that parameters like surface gravity and mass derived from spectral fitting are generally more accurate and reliable. Moreover, derived $T_{\rm eff}$ from $Gaia$ are much more consistent with LAMOST $T_{\rm eff}$ when they are low. This implies derived $T_{\rm eff}$ from $Gaia$ for relatively cool WDs are generally reliable, while for hotter WDs, derived $T_{\rm eff}$ should be used cautiously.

With respect to the parameter distributions of the 70 relatively high S/N DBs, the peak in mass is located near 0.65\,M$_{\odot}$, and the peak in log\,$g$ distribution is around 8.0. These distributions are consistent with the literature \citep{Eisenstein2006,Kleinman2013}. It is difficult to draw any further conclusions from these distributions, because there are so few sources.

\begin{figure}
\center
\includegraphics[angle=0,width=0.5\textwidth]{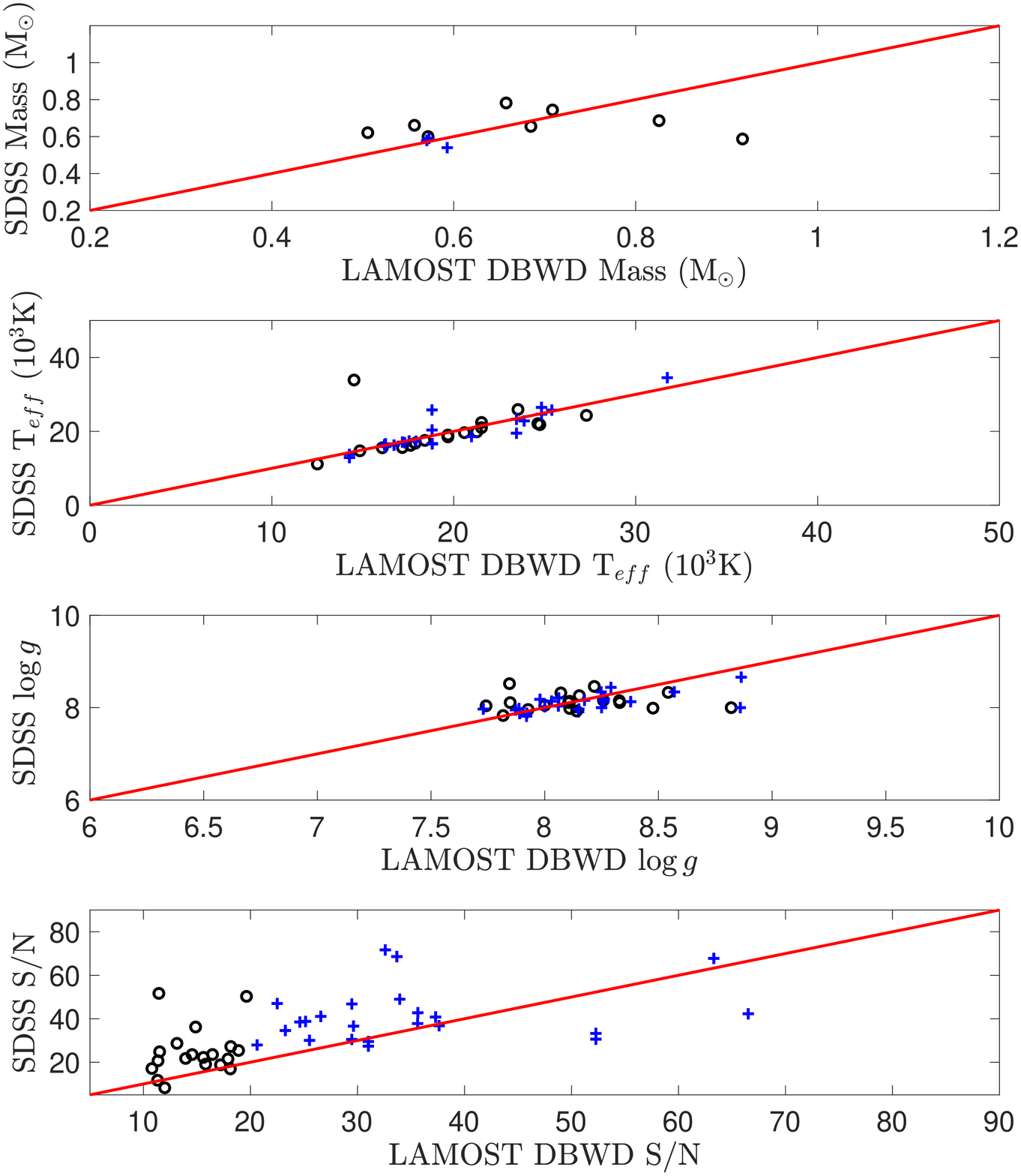}
\caption{Comparisons between LAMOST DB parameters and SDSS parameters from \protect\cite{Kleinman2013} and \protect\cite{Koester2015}. From top to bottom: mass, effective temperature, surface gravity and S/N. Black circles represent DBs with S/N of LAMOST spectra between 10 and 20, while blue crosses are DAs with S/N of LAMOST spectra greater than 20. The red solid lines are unit slope relation.}
\label{fig7}
\end{figure}

Among the 70 DBs with derived parameters, 42 have parameters in \cite{Kleinman2013} and \cite{Koester2015} , but only 11 sources have derived masses. In Fig. \ref{fig7}, from top to bottom panel, these are comparisons of mass, effective temperature, surface gravity and S/N between our results and results from \cite{Kleinman2013} and \cite{Koester2015}, respectively. Based on these direct comparisons, our results are generally consistent with results from SDSS. It is also true that DBs with higher S/N tend to have more consistent parameters than DBs with lower S/N. And SDSS DB spectra have higher S/N than SDSS spectra, when the S/N of LAMOST DBs are less than $\sim$ 50. Even though there are only three sources, it's likely the opposite when the S/N of LAMOST DBs are greater than $\sim$ 50.

\section{On going and future WD candidate selection for LAMOST}

We have been selecting potential WD targets for future, low-resolution, LAMOST spectral fibre selection. In this section, we describe our pre-{\em Gaia} DR2 criteria for the second phase survey. Even though with the release of {\em Gaia} 2$^{\rm nd}$ data release, a white dwarf candidates sample as large as 486\,641 has been revealed \citep{Gentile2018}, and their $T_{\rm eff}$, log\,$g$, radii and mass can be inferred from accurate photometry and trigonometric parallax. However, the importance of obtaining the optical white dwarf spectra can not be neglected. One important application is to discover more rare WDs with emission lines caused by gaseous debris disk or irradiation from unseen substellar object \citep{Gansicke2006,Melis2012,Guo2015a,Manser2019}. Another example is to identify more metal polluted WDs (DZs), which can be a unique tool to analyse the chemical composition of their tidally disrupted planetesimal \citep{Gansicke2012,Farihi2013}. As a matter of fact, in order to obtain WD optical spectra, future large multi-object spectroscopic surveys are planning to observe these white dwarf candidates from {\em Gaia}. For instance, SDSS-V \citep{Kollmeier2017}, 4most \citep{de Jong2012}, WEAVE \citep{Dalton2012}, and DESI \citep{DESI2016}.

\begin{figure}
\center
\includegraphics[angle=90,width=0.5\textwidth]{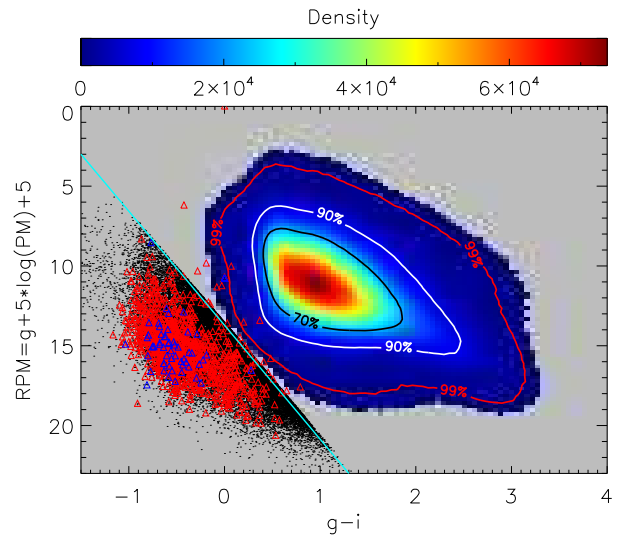}
\caption{Colour g-i versus RPM. The coloured density map and contours demonstrate the majority of common stars locate at the central region. Red triangles represent known DAs in the common source sample, while blue triangles represent known DBs. The cyan solid line is defined to separate WDs from majority of stars. The black dots are selected white dwarf candidates, based on the RPM approach.}
\label{fig8}
\end{figure}

At the end of the LAMOST first phase survey in 2017 June, we began a process of selecting WD candidates, based on reduced proper motion. The photometric data we used is taken from the Xuyi Schmidt Telescope Photometric Survey of the Galactic Anti-center \citep[XSTPS-GAC]{LiuXW2014,Guo2018} and Pan-STARRS \citep{Kaiser2010}, while the proper motion data are from the Gaia-PS1-SDSS \citep[GPS1]{Tian2017} proper motion catalog. XSTPS-GAC is a photometric sky survey in the SDSS $gri$ bands, and because LAMOST sources are selected from XSTPS-GAC, it is highly suitable for WD candidate identification. The total number of sources in XSTPS-GAC is 110\,168\,720, and after cross-matching with GPS1 within 3\,arcsec, there are 56\,059\,242 sources in common. It is worth mentioning that this is a simple direct cross-matching without considering the epochs of observations and proper motion of stars. We use equation \ref{equation1} to calculate reduced proper motion, and plot this against $g-i$ in order to identify WD candidates, where Fig.\ \ref{fig8} shows our exact parameter cuts.

Common sources are plotted in the $g$-$i$ colour versus RPM space, over-plotted with the density contours. Red triangles represent 2\,030 known DAs among these common sources, while blue triangles represent 127 known DBs in the same sample. Based on their location, WDs should locate in the same region. Therefore, we defined a triangle region to separate the WDs and the other types of star. The triangle region is defined by cyan solid line, X axis and Y axis. Black dots are common sources locate in the defined region, which represent selected white dwarf candidates. 

Our reduced proper motion cuts yielded 30\,441 WD candidates. Next, we cross-matched these objects with the literature, finding 20\,628 sources are previously unrecognised WD candidates (at that time). These candidates were entered into the target selection software designed for LAMOST. Some of those white dwarf candidates have already been observed and included in DR6, while the rest will appear in the 2nd phase.

\begin{equation}
RPM=g+5\times log(PM)+5
\label{equation1}
\end{equation}

After the second release of {\em Gaia} data in 2018 April, \cite{Gentile2018} has revealed a white dwarf candidate sample as large as 486\,k. Unfortunately, our selection of WDs for LAMOST second phase has completed in 2017. Nonetheless, plan has been made to utilise this high fidelity catalog to form a proper sample for LAMOST to observe in the future.

\section{Summary}
In this work, we study the DAs and DBs identified in LAMOST DR5. The LAMOST pipeline classification, colour-colour cut and random forest method were utilised to select DA candidates. Visual inspection produced 2\,620 authentic DA spectra. For DBs, a random forest, machine learning algorithm was used to select the candidates. From 12\,572 DB candidates, 182 were visually confirmed. Regarding the efficiency, larger training sample in the future can definitely be helpful. After cross-matching with SIMBAD and the literature, 393 DAs and 46 DBs are found to be new identifications. 

Those identified WDs were analysed. For all DAs where their spectral S/N $>10$, and DBs with same S/N $>10$, their spectra were fitted with white dwarf atmosphere models to derive effective temperature and surface gravity \citep{Koester2010}. From these two parameters the mass and cooling age were calculated from evolutionary models. 

Proper motions, magnitude and parallax of {\em Gaia} DR2 are also provided in our catalog (Table. \ref{tab3}). In addition, distributions of mass, effective temperature, surface gravity, cooling age and distance were built to study the property of our sample.

For DAs, the peak of the mass distribution is in general agreement with previous results. Both low- and high-mass residuals are present in the mass distributions. For DBs, the peak of the mass distribution is located around 0.65 M$_{\odot}$, and also consistent with other studies. However, there are only 70 DBs with sufficiently high S/N to derive reliable parameters, and the resulting distributions are too sparsely populated to draw any conclusions. The parameters of LAMOST DAs and DBs have been compared for sources in common with previous work, and found to be in good agreement for spectra with relatively high S/N. We also found that for spectra with LAMOST S/N less than $\sim$ 50, SDSS spectra tend to have larger S/N, while for spectra with S/N greater than $\sim$ 50, LAMOST spectra have larger S/N.

Even after the release of {\em Gaia} DR2 \citep{Gaia2018} and the white dwarf candidates catalog in {\em Gaia} \citep{Gentile2018}, the observation of optical white dwarf spectra is still of great importance. 

Optical spectral identification is of increasing importance in the {\em Gaia} era. Therefore, we selected  20\,628 white dwarf candidates from XSTPS-GAC and GPS1, then put them into the input catalog of low-resolution observation for the transition year and 2$^{\rm nd}$ phase of LAMOST, which began in 2017. Since the white dwarf candidates catalog from {\em Gaia} DR2 has been published, we will try to add those observable candidates into LAMOST inout catalog in the near future as well.

\section*{Acknowledgments}
We thank the anonymous referee for helpful suggestions and advices, which improved this paper a lot. We thank M.Barlow, J.Farihi and A.Swan for a careful reading of the manuscript and helping with the improvement of English writing. We also thank D. Koester and P. Bergeron for providing white dwarf models. Balmer/Lyman lines in the models were calculated with the modified Stark broadening profiles of \cite{Tremblay2009}, kindly made available by the authors. This work is supported by the National Natural Science Foundation of China (Grant No. 11890694, No. 11473001 and 11078006). The authors also acknowledge support by the National Key Basic Research Program of China (Grant No. 2014CB845700). We also acknowledge the support from the 2m Chinese Space Station Telescope project: CMS-CSST-2021-A10. This work was partially supported by the Scholar Program of Beijing Academy of Science and Technology (DZ:BS202002). This work is also supported by the China Postdoctoral Science Foundation (Grant No. 2017M610695) and the Astronomical Big Data Joint Research Centre, co-founded by the National Astronomical Observatories, Chinese Academy of Sciences and the Alibaba Cloud.

The LAMOST FELLOWSHIP is supported by Special Funding for Advanced Users, budgeted and administrated by the Centre for Astronomical Mega-Science, Chinese Academy of Sciences.

The Guo Shou Jing Telescope (the Large Sky Area Multi-Object Fibre Spectroscopic Telescope, LAMOST) is a National Major Scientific Project which is built by the Chinese Academy of Sciences, funded by the National Development and Reform Commission, and operated and
managed by the National Astronomical Observatories, Chinese Academy of Sciences.

\section*{DATA AVAILABILITY}
The spectra data underlying this article are publicly available from LAMOST archive (\url{http://dr5.lamost.org/}). Table \ref{tab2}, \ref{tab3}, \ref{tab5}, and \ref{tab6} are partially shown in the article. They are fully accessible through publisher in supplementary material. Table \ref{tab5} and \ref{tab6} are WDs identified in our previous work (LAMOST DR2) with updated classifications and parameters.


\newpage

\begin{table*}
\caption{
Basic parameters of WDs from LAMOST DR 3, 4 \& 5. \\
(Only part of the catalogue is shown here. The full table can be found in the supplementary material.)
}
\begin{center}
\begin{threeparttable}
\label{tab2}
{\scriptsize
\begin{tabular}{p{0.3cm} c c c c c c c c c c }\hline
\hline
GID$^{a}$ & ObsID$^{b}$ & Designation$^{c}$ & RA & DEC& Obs-date$^{d}$ & Mjd-Planid$\_$spid-Fiberid$^{e}$ & Type \\
	           &	      		&					      &			  &          &		    &  &  & \\\hline

1$\_$2 & 247515227 & J002633.24+390902.9 & 6.638503 & 39.150817 & 2014-09-10 & 56911-HD002951N381926B01$\_$sp15-227 & DAZ\\
1$\_$2 & 354311163 & J002633.13+390904.0 & 6.6380824 & 39.151121 & 2015-09-14 & 57280-M31007N36M1$\_$sp11-163 & DAZ\\
2$\_$2 & 249301140 & J041010.37+180222.8 & 62.543227 & 18.039691 & 2014-09-26 & 56927-GAC062N19B1$\_$sp01-140 & DA\\
2$\_$2 & 318401139 & J041010.37+180222.8 & 62.543227 & 18.039691 & 2015-02-13 & 57067-GAC062N19B2$\_$sp01-139 & DA\\
3$\_$2 & 250301120 & J042355.72+162113.2 & 65.982207 & 16.353693 & 2014-10-05 & 56936-GAC065N18B1$\_$sp01-120 & DA\\
3$\_$2 & 250401120 & J042355.78+162114.9 & 65.982452 & 16.354149 & 2014-10-05 & 56936-GAC065N18B2$\_$sp01-120 & DA\\
4$\_$2 & 250407081 & J042839.47+165811.7 & 67.164496 & 16.969927 & 2014-10-05 & 56936-GAC065N18B2$\_$sp07-081 & DA\\
4$\_$2 & 420109239 & J042839.40+165812.0 & 67.1642003 & 16.9700186 & 2016-02-05 & 57424-KP042325N164638B01$\_$sp09-239 & DA\\
5$\_$2 & 252902097 & J024746.29+000331.6 & 41.942909 & 0.0587914 & 2014-10-13 & 56944-EG025338N015809M01$\_$sp02-097 & DA\\
5$\_$2 & 367416195 & J024746.39+000331.1 & 41.943312 & 0.058658 & 2015-10-08 & 57304-EG025335S013827B01$\_$sp16-195 & DA\\
6$\_$2 & 252908132 & J025817.87+010946.0 & 44.574461 & 1.1627782 & 2014-10-13 & 56944-EG025338N015809M01$\_$sp08-132 & DA+M\\
6$\_$2 & 300710119 & J025817.87+010946.0 & 44.574461 & 1.1627782 & 2015-01-20 & 57043-EG030739N012421M01$\_$sp10-119 & DA+M\\
7$\_$3 & 254115228 & J040342.10+145928.7 & 60.9254278 & 14.9913211 & 2014-10-15 & 56946-HD040531N141710B01$\_$sp15-228 & DA\\
7$\_$3 & 381910034 & J040342.21+145929.8 & 60.925876 & 14.991617 & 2015-11-03 & 57330-GAC062N15B1$\_$sp10-034 & DA\\
7$\_$3 & 382010034 & J040342.19+145928.4 & 60.925814 & 14.991239 & 2015-11-03 & 57330-GAC062N15B2$\_$sp10-034 & DA\\
8$\_$2 & 256106072 & J225745.89+074320.5 & 344.44124 & 7.722373 & 2014-10-25 & 56956-EG224840N075042M01$\_$sp06-072 & DA\\
8$\_$2 & 380401098 & J225745.89+074320.5 & 344.44124 & 7.722373 & 2015-11-01 & 57328-EG225829N094931M01$\_$sp01-098 & DA\\
9$\_$2 & 256113149 & J225605.44+081936.3 & 344.0227 & 8.32677 & 2014-10-25 & 56956-EG224840N075042M01$\_$sp13-149 & DA\\
9$\_$2 & 380402065 & J225605.44+081936.3 & 344.0227 & 8.32677 & 2015-11-01 & 57328-EG225829N094931M01$\_$sp02-065 & DA\\
10$\_$2 & 256201054 & J055944.97+171203.3 & 89.937392 & 17.20094 & 2014-10-25 & 56956-GAC090N18M1$\_$sp01-054 & DA\\
10$\_$2 & 420409046 & J055944.97+171203.4 & 89.937412 & 17.200946 & 2016-02-06 & 57425-GAC089N16M1$\_$sp09-046 & DA\\
11$\_$2 & 256601015 & J061518.94+153059.4 & 93.828958 & 15.5165 & 2014-10-26 & 56957-GAC093N17M1$\_$sp01-015 & CV\\
11$\_$2 & 400112059 & J061518.94+153059.4 & 93.828958 & 15.5165 & 2016-01-04 & 57392-GAC092N13M1$\_$sp12-059 & CV\\
12$\_$2 & 257606145 & J071004.93+292403.2 & 107.52056 & 29.400907 & 2014-11-01 & 56963-GAC105N29B1$\_$sp06-145 & DA\\
12$\_$2 & 281116178 & J071004.93+292403.2 & 107.52056 & 29.400907 & 2014-12-18 & 57010-GAC108N27B1$\_$sp16-178 & DA\\
13$\_$3 & 259201037 & J075853.02+161645.1 & 119.72092 & 16.279214 & 2014-11-04 & 56966-GAC120N18B1$\_$sp01-037 & CV\\
13$\_$3 & 392716175 & J075853.02+161645.1 & 119.72092 & 16.279214 & 2015-12-21 & 57378-GAC120N14B1$\_$sp16-175 & CV\\
13$\_$3 & 392816175 & J075853.02+161645.1 & 119.72092 & 16.279214 & 2015-12-21 & 57378-GAC120N14B2$\_$sp16-175 & CV\\
14$\_$2 & 265114068 & J060911.91+352549.3 & 92.299661 & 35.430376 & 2014-11-12 & 56974-GAC094N35B1$\_$sp14-068 & DA\\
14$\_$2 & 385314070 & J060912.00+352549.1 & 92.300011 & 35.430329 & 2015-12-01 & 57358-GAC094N35B2$\_$sp14-070 & DA\\
15$\_$2 & 266807242 & J022301.66+061649.5 & 35.756946 & 6.2804389 & 2014-11-16 & 56978-EG021841N081050M01$\_$sp07-242 & CV\\
15$\_$2 & 474703167 & J022301.66+061649.5 & 35.756946 & 6.280439 & 2016-11-02 & 57695-EG022616N061733M01$\_$sp03-167 & CV\\
\hline

\end{tabular}
\begin{tablenotes}
\item Notes: \newline
$^{a}$Group ID. i.e. ``1\_2" means group 1 has 2 spectra.\newline
$^{b}$Observation ID, which is also unique for each spectrum.\newline
$^{c}$Designation from LAMOST. The exact number could be slight different for the same group.\newline
$^{d}$Observation date of the spectrum. \newline
$^{e}$Unique spectrum ID. \newline
\end{tablenotes}
}

\end{threeparttable}
\end{center}

\end{table*}

\begin{table*}
\caption{Determined WD parameters and parameters from Gaia DR2. \newline
(Only part of the catalogue is shown here. The full table can be found in the supplementary material.)
}
\begin{center}
\begin{threeparttable}[hb]

\label{tab3}
{\scriptsize
\begin{tabular}{p{0.3cm} c c c c c c c c c c c}\hline
\hline
GID$^{a}$ & ObsID$^{b}$ & S/N$_{g}$$^{c}$& $T_{\rm eff}$ & log\,$g$ & Mass & Age &  Plx$^{d}$ & PMra$^{e}$ & PMdec$^{e}$ & mag$_{G}$$^{f}$ &Type \\
	           &	      		&		  &  (K)          &		    & (M$_{\odot}$)& (Myr)& (mas) & (mas/yr) & (mas/yr) & & \\\hline

1$\_$2 & 247515227 & 11 &  &  & & & 15.8342$\pm$0.066 & 142.505$\pm$0.102 & -130.002$\pm$0.087 & 15.800 & DAZ \\
1$\_$2 & 354311163 & 3 &  &  & & & 15.8342$\pm$0.066 & 142.505$\pm$0.102 & -130.002$\pm$0.087 & 15.800 & DAZ \\
2$\_$2 & 249301140 & 105 &  &  & & & 29.3892$\pm$0.0572 & 72.19$\pm$0.14 & -85.906$\pm$0.096 & 14.244 & DA \\
2$\_$2 & 318401139 & 124 & 15312$\pm$1369 & 8.04$\pm$0.27 & 0.64$\pm$0.16 & 199$\pm$114 & 29.3892$\pm$0.0572 & 72.19$\pm$0.14 & -85.906$\pm$0.096 & 14.244 & DA \\
3$\_$2 & 250301120 & 22 &  &  & & & 22.2272$\pm$0.0519 & 114.412$\pm$0.103 & -27.715$\pm$0.079 & 14.347 & DA \\
3$\_$2 & 250401120 & 39 & 20526$\pm$1688 & 8.24$\pm$0.27 & 0.77$\pm$0.17 & 109$\pm$73 & 22.2272$\pm$0.0519 & 114.412$\pm$0.103 & -27.715$\pm$0.079 & 14.347 & DA \\
4$\_$2 & 250407081 & 115 &  &  & & & 20.8952$\pm$0.0567 & 102.692$\pm$0.115 & -26.885$\pm$0.068 & 14.074 & DA \\
4$\_$2 & 420109239 & 116 & 26029$\pm$1851 & 8.14$\pm$0.29 & 0.72$\pm$0.18 & 27$\pm$4 & 20.8952$\pm$0.0567 & 102.692$\pm$0.115 & -26.885$\pm$0.068 & 14.074 & DA \\
5$\_$2 & 252902097 & 9 &  &  & & & 8.1485$\pm$0.0776 & 130.453$\pm$0.136 & -41.637$\pm$0.125 & 16.488 & DA \\
5$\_$2 & 367416195 & 35 & 19475$\pm$1519 & 8.07$\pm$0.27 & 0.66$\pm$0.16 & 92$\pm$65 & 8.1485$\pm$0.0776 & 130.453$\pm$0.136 & -41.637$\pm$0.125 & 16.488 & DA \\
6$\_$2 & 252908132 & 7 &  &  & & &  &  &  & 17.818 & DA+M \\
6$\_$2 & 300710119 & 5 &  &  & & &  &  &  & 17.818 & DA+M \\
7$\_$3 & 254115228 & 101 & 14421$\pm$2543 & 8.43$\pm$0.27 & 0.88$\pm$0.17 & 442$\pm$283 & 24.0527$\pm$0.0541 & 141.19$\pm$0.107 & -24.07$\pm$0.084 & 15.038 & DA \\
7$\_$3 & 381910034 & 53 &  &  & & & 24.0527$\pm$0.0541 & 141.19$\pm$0.107 & -24.07$\pm$0.084 & 15.038 & DA \\
7$\_$3 & 382010034 & 42 &  &  & & & 24.0527$\pm$0.0541 & 141.19$\pm$0.107 & -24.07$\pm$0.084 & 15.038 & DA \\
8$\_$2 & 256106072 & 7 &  &  & & & 4.557$\pm$0.2331 & -15.071$\pm$0.374 & -40.83$\pm$0.269 & 18.302 & DA \\
8$\_$2 & 380401098 & 12 &  &  & & & 4.557$\pm$0.2331 & -15.071$\pm$0.374 & -40.83$\pm$0.269 & 18.302 & DA \\
9$\_$2 & 256113149 & 6 &  &  & & & 6.3772$\pm$0.1 & -19.921$\pm$0.199 & -1.312$\pm$0.136 & 16.738 & DA \\
9$\_$2 & 380402065 & 37 & 23254$\pm$2294 & 7.65$\pm$0.34 & 0.47$\pm$0.13 & 20$\pm$7 & 6.3772$\pm$0.1 & -19.921$\pm$0.199 & -1.312$\pm$0.136 & 16.738 & DA \\
10$\_$2 & 256201054 & 57 &  &  & & & 12.713$\pm$0.0718 & 0.726$\pm$0.121 & -19.04$\pm$0.1 & 15.841 & DA \\
10$\_$2 & 420409046 & 87 & 18982$\pm$1348 & 8.31$\pm$0.25 & 0.81$\pm$0.16 & 162$\pm$98 & 12.713$\pm$0.0718 & 0.726$\pm$0.121 & -19.04$\pm$0.1 & 15.841 & DA \\
11$\_$2 & 256601015 & 4 &  &  & & & 1.2902$\pm$0.1754 & -0.069$\pm$0.31 & -8.973$\pm$0.311 & 17.961 & CV \\
11$\_$2 & 400112059 & 3 &  &  & & & 1.2902$\pm$0.1754 & -0.069$\pm$0.31 & -8.973$\pm$0.311 & 17.961& CV \\
12$\_$2 & 257606145 & 19 &  &  & & & 14.4462$\pm$0.0781 & -1.383$\pm$0.114 & -101.355$\pm$0.102 & 15.515 & DA \\
12$\_$2 & 281116178 & 28 & 15710$\pm$1559 & 7.97$\pm$0.31 & 0.60$\pm$0.17 & 162$\pm$95 & 14.4462$\pm$0.0781 & -1.383$\pm$0.114 & -101.355$\pm$0.102 & 15.515 & DA \\
13$\_$3 & 259201037 & 43 &  &  & & & 4.79$\pm$0.0572 & -26.015$\pm$0.116 & -0.264$\pm$0.063 & 15.515 & CV \\
13$\_$3 & 392716175 & 26 &  &  & & & 4.79$\pm$0.0572 & -26.015$\pm$0.116 & -0.264$\pm$0.063 & 15.515 & CV \\
13$\_$3 & 392816175 & 41 &  &  & & & 4.79$\pm$0.0572 & -26.015$\pm$0.116 & -0.264$\pm$0.063 & 15.515 & CV \\
14$\_$2 & 265114068 & 16 &  &  & & & 20.6106$\pm$0.0851 & -0.299$\pm$0.107 & 51.212$\pm$0.096 & 15.681 & DA \\
14$\_$2 & 385314070 & 33 & 10214$\pm$360 & 8.39$\pm$0.35 & 0.85$\pm$0.23 & 1064$\pm$711 & 20.6106$\pm$0.0851 & -0.299$\pm$0.107 & 51.212$\pm$0.096 & 15.681 & DA \\
15$\_$2 & 266807242 & 34 &  &  & & & 1.3651$\pm$0.1253 & 5.159$\pm$0.164 & -8.974$\pm$0.154 & 17.051 & CV \\
15$\_$2 & 474703167 & 43 &  &  & & & 1.3651$\pm$0.1253 & 5.159$\pm$0.164 & -8.974$\pm$0.154 & 17.051 & CV \\

\hline

\end{tabular}
\begin{tablenotes}
\item Notes: \newline
$^{a}$Group ID. i.e. ``1\_2" means group 1 has 2 spectra.\newline
$^{b}$Unique ID for each spectrum.\newline
$^{c}$Signal to Noise ratio in g band. Note only spectra with S/N>=15 in g band are fitted.\newline
$^{d}$Parallax from {\em Gaia} DR2 in mas.\newline
$^{e}$Proper motions in ra and dec from Gaia DR2 in units of mas/yr.\newline
$^{f}$G band photometry from Gaia DR2. \newline
\end{tablenotes}
}

\end{threeparttable}
\end{center}

\end{table*}

\begin{table*}
\caption{Mis-classified white dwarfs in LAMOST DR2. \newline
(Only part of the catalogue is shown here. The full table can be found in the supplementary material.)
}
\begin{center}
\begin{threeparttable}[hb]

\label{tab5}
{\scriptsize
\begin{tabular}{p{0.3cm} c c c c c c c c c c }\hline
\hline
GID$^{a}$  & RA & DEC& Designation$^{b}$ & Mjd-Planid$\_$spid-Fiberid$^{c}$ & snrg &  Old type & Update\\
	          &	  &          &		                      &                                                       &        &           &             \\\hline

64$\_$2 & 98.8159 & 2.39865 & J063515.81+022355.1 & 55976-GAC$\_$099N04$\_$V1$\_$sp02-124 & 11.73 & DA$\_$in$\_$nebula? & star\\ 
64$\_$2 & 98.8159 & 2.39865 & J063515.81+022355.1 & 55976-GAC$\_$099N04$\_$V5$\_$sp02-124 & 10.78 & DA$\_$in$\_$nebula? & star\\ 
111$\_$3 & 106.68858 & 22.26037 & J070645.25+221537.3 & 56611-GAC108N21V3$\_$sp14-199 & 28.55 & DA+M? & star\\ 
111$\_$3 & 106.68858 & 22.26037 & J070645.25+221537.3 & 56621-GAC108N21V1$\_$sp14-199 & 11.85 & DA+M? & star\\ 
111$\_$3 & 106.68858 & 22.26037 & J070645.25+221537.3 & 56621-GAC108N21V3$\_$sp14-199 & 15.67 & DA+M? & star\\ 
113$\_$2 & 58.28319 & 34.53968 & J035307.96+343222.8 & 56618-VB057N34V1$\_$sp08-238 & 17.45 & DAZ & star\\ 
113$\_$2 & 58.28319 & 34.53968 & J035307.96+343222.8 & 56618-VB057N34V2$\_$sp08-238 & 10.69 & DAZ & star\\ 
141$\_$1 & 80.71046 & 29.65543 & J052250.50+293919.5 & 55859-M5904$\_$sp15-198 & 3.41 & DA? & star\\ 
142$\_$1 & 10.58862 & 39.37988 & J004221.26+392247.5 & 55862-M6201$\_$sp16-091 & 5.67 & DA? & star\\ 
143$\_$1 & 12.11577 & 39.62563 & J004827.78+393732.2 & 55862-M6201$\_$sp16-122 & 5.27 & DA? & star\\ 
151$\_$1 & 89.49879 & 29.20391 & J055759.71+291214.0 & 55876-GAC$\_$089N28$\_$B2$\_$sp04-060 & 5.97 & DA? & star\\ 
152$\_$1 & 89.49431 & 29.98647 & J055758.63+295911.2 & 55876-GAC$\_$089N28$\_$B2$\_$sp15-134 & 3.53 & DA? & star\\ 
153$\_$1 & 91.57595 & 28.03239 & J060618.22+280156.5 & 55876-GAC$\_$089N28$\_$B3$\_$sp06-113 & 4.95 & DA? & star\\ 
154$\_$1 & 89.50101 & 29.54909 & J055800.24+293256.7 & 55876-GAC$\_$089N28$\_$B3$\_$sp15-102 & 7.67 & DA? & star\\ 
156$\_$1 & 88.77986 & 29.0578 & J055507.16+290328.0 & 55876-GAC$\_$089N28$\_$B1$\_$sp04-001 & 5.23 & DA? & star\\ 
158$\_$1 & 53.9205 & 4.24866 & J033540.91+041455.1 & 55886-F8606$\_$sp05-206 & 5.68 & DA? & star\\ 
170$\_$1 & 71.51261 & 26.14234 & J044603.02+260832.4 & 55890-GAC$\_$067N28$\_$M1$\_$sp01-236 & 5.18 & DA? & star\\ 
171$\_$1 & 74.26587 & 27.71918 & J045703.80+274309.0 & 55890-GAC$\_$067N28$\_$M1$\_$sp06-197 & 6.43 & DA? & star\\ 
173$\_$1 & 106.81713 & 26.6493 & J070716.11+263857.4 & 55890-GAC$\_$106N28$\_$M1$\_$sp01-154 & 9.85 & DA+M? & DA\\ 
182$\_$1 & 50.2794 & 4.86407 & J032107.05+045150.6 & 55892-F9204$\_$sp04-169 & 8.39 & DBA? & hotSD\\

\hline

\end{tabular}
\begin{tablenotes}
\item Notes: \newline
$^{a}$Group ID. i.e. ``1\_2" means group 1 has 2 spectra.\newline
$^{b}$Designation from LAMOST. The exact number could be slight different for the same group.\newline
$^{c}$Unique spectrum ID. \newline
\end{tablenotes}
}

\end{threeparttable}
\end{center}

\end{table*}

\begin{table*}
\caption{DA White dwarfs with updated parameters in LAMOST DR2. \newline
(Only part of the catalogue is shown here. The full table can be found in the supplementary material.)
}
\begin{center}
\begin{threeparttable}[hb]

\label{tab6}
{\scriptsize
\begin{tabular}{p{0.5cm} c c c c c c c c c c }\hline
\hline
ObsID$^{a}$ & Designation$^{b}$ & Obsdate & RA& DEC & snrg & Teff$^{c}$  & logg$^{c}$ & mass$^{c}$ & age$^{c}$ \\ 
	          &	  &          &		                      &                                                       &        &           &             \\\hline

112211 & J221640.39+012741.2 & 2011-10-24 & 334.168324 & 1.461445 & 10.49 & 10872$\pm$360 & 8.28$\pm$0.36 & 0.78$\pm$0.23 & 740$\pm$400\\ 
116148 & J220522.86+021837.5 & 2011-10-24 & 331.345250 & 2.310432 & 14.26 & 13000$\pm$2793 & 7.76$\pm$0.35 & 0.48$\pm$0.16 & 218$\pm$167\\ 
403199 & J004128.66+402324.0 & 2011-10-24 & 10.369458 & 40.390026 & 19.99 & 27493$\pm$2182 & 7.60$\pm$0.35 & 0.46$\pm$0.13 & 11$\pm$3\\ 
1102134 & J025737.24+264047.8 & 2011-10-27 & 44.405201 & 26.679970 & 15.86 & 10838$\pm$359 & 8.51$\pm$0.33 & 0.93$\pm$0.21 & 1134$\pm$1078\\ 
1115183 & J030214.72+285707.4 & 2011-10-27 & 45.561340 & 28.952057 & 13.54 & 16814$\pm$1427 & 7.53$\pm$0.31 & 0.39$\pm$0.10 & 66$\pm$23\\ 
1506145 & J071004.83+292402.8 & 2011-10-28 & 107.520140 & 29.400780 & 12.06 & 12000$\pm$3141 & 6.88$\pm$0.50 & 0.21$\pm$0.10 & 113$\pm$147\\ 
1615243 & J004036.79+413138.7 & 2011-10-28 & 10.153296 & 41.527443 & 21.71 & 13564$\pm$2707 & 7.83$\pm$0.32 & 0.52$\pm$0.17 & 209$\pm$183\\ 
1616129 & J003956.55+422929.5 & 2011-10-28 & 9.985629 & 42.491542 & 10.24 & 9861$\pm$303 & 7.76$\pm$0.55 & 0.47$\pm$0.26 & 472$\pm$266\\ 
2101118 & J005340.52+360116.8 & 2011-11-08 & 13.418857 & 36.021358 & 32.82 & 28460$\pm$1577 & 7.85$\pm$0.33 & 0.57$\pm$0.15 & 10$\pm$1\\ 
2102081 & J004159.19+362352.3 & 2011-11-08 & 10.496630 & 36.397865 & 10.03 & 22062$\pm$2314 & 7.60$\pm$0.33 & 0.44$\pm$0.13 & 24$\pm$9\\ 
6610008 & J032942.79+053755.8 & 2011-11-20 & 52.428301 & 5.632185 & 11.08 & 19155$\pm$1518 & 8.04$\pm$0.27 & 0.65$\pm$0.16 & 91$\pm$67\\ 
6807221 & J063532.48+261958.6 & 2011-11-20 & 98.885366 & 26.332951 & 14.00 & 35843$\pm$2393 & 8.00$\pm$0.42 & 0.67$\pm$0.21 & 5$\pm$2\\ 
6904155 & J080800.00+290152.5 & 2011-11-20 & 122.000010 & 29.031273 & 19.48 & 25340$\pm$2300 & 7.93$\pm$0.32 & 0.60$\pm$0.15 & 17$\pm$8\\ 
7015243 & J004036.79+413138.7 & 2011-11-20 & 10.153296 & 41.527443 & 19.41 & 13000$\pm$2414 & 7.86$\pm$0.35 & 0.53$\pm$0.19 & 249$\pm$201\\ 
7516178 & J071004.83+292402.8 & 2011-11-23 & 107.520140 & 29.400789 & 12.04 & 13758$\pm$1742 & 8.16$\pm$0.30 & 0.71$\pm$0.19 & 324$\pm$226\\ 
7801231 & J100551.51-023417.8 & 2011-11-24 & 151.464628 & -2.571630 & 19.69 & 20835$\pm$1816 & 7.99$\pm$0.30 & 0.62$\pm$0.17 & 55$\pm$58\\

\hline

\end{tabular}
\begin{tablenotes}
\item Notes: \newline
$^{a}$Unique ID for each spectrum.\newline
$^{b}$Designation from LAMOST. The exact number could be slight different for the same group.\newline
$^{c}$Undated parameters using fitting method adopted in this work.\newline
\end{tablenotes}
}

\end{threeparttable}
\end{center}

\end{table*}


\begin{thebibliography}{99}

\bibitem[\protect\citeauthoryear{Abazajian et al.}{2009}]{Abazajian2009} Abazajian K.~N., et al., 2009, ApJS, 182, 543 


\bibitem[\protect\citeauthoryear{Ahn et al.}{2014}]{Ahn2014} Ahn C.~P., et al., 2014, ApJS, 211, 17 


\bibitem[\protect\citeauthoryear{Bai, Liu, \& Wang}{2018}]{Bai2018} Bai Y., Liu J.-F., Wang S., 2018, RAA, 18, 118 

\bibitem[\protect\citeauthoryear{Bai et al.}{2019}]{Bai2019} Bai Y., Liu J., Wang S., Yang F., 2019, AJ, 157, 9 


\bibitem[\protect\citeauthoryear{Bergeron, Leggett, \& Ruiz}{2001}]{Bergeron2001} Bergeron P., Leggett S.~K., Ruiz M.~T., 2001, ApJS, 133, 413 

\bibitem[\protect\citeauthoryear{Bovy}{2017}]{Bovy2017} Bovy J., 2017, MNRAS, 468, L63


\bibitem[\protect\citeauthoryear{Breiman}{2001}]{Breiman2001} Breiman. L., 2001, Machine Learning, 45, 5 

\bibitem[\protect\citeauthoryear{Carlin et al.}{2012}]{Carlin2012} Carlin J.~L., et al., 2012, RAA, 12, 755 


\bibitem[\protect\citeauthoryear{Catal{\'a}n et al.}{2008a}]{Catalan2008} Catal{\'a}n S., Isern J., Garc{\'{\i}}a-Berro E., Ribas I., 2008, MNRAS, 387, 1693 

\bibitem[\protect\citeauthoryear{Catal{\'a}n et al.}{2008b}]{Catalan2008b} Catal{\'a}n S., Isern J., Garc{\'\i}a-Berro E., Ribas I., Allende Prieto C., Bonanos A.~Z., 2008, A\&A, 477, 213

\bibitem[\protect\citeauthoryear{Chen et al.}{2012}]{Chen2012} Chen L., et al., 2012, RAA, 12, 805 


\bibitem[\protect\citeauthoryear{Cui et al.}{2012}]{Cui2012} Cui X.-Q., et al., 2012, RAA, 12, 1197 

\bibitem[\protect\citeauthoryear{Cunningham et al.}{2020}]{Cunningham2020} Cunningham T., Tremblay P.-E., Gentile Fusillo N.~P., Hollands M., Cukanovaite E., 2020, MNRAS, 492, 3540. 

\bibitem[\protect\citeauthoryear{Dalton et al.}{2012}]{Dalton2012} Dalton G., et al., 2012, SPIE, 8446, 84460P 


\bibitem[\protect\citeauthoryear{de Jong et al.}{2012}]{de Jong2012} de Jong R.~S., et al., 2012, SPIE, 8446, 84460T 


\bibitem[\protect\citeauthoryear{Deng et al.}{2012}]{Deng2012} Deng L.-C., et al., 2012, RAA, 12, 735 


\bibitem[\protect\citeauthoryear{DESI Collaboration et al.}{2016}]{DESI2016} DESI Collaboration, et al., 2016, arXiv, arXiv:1611.00036 

\bibitem[\protect\citeauthoryear{Doherty et al.}{2015}]{Doherty2015} Doherty C.~L., Gil-Pons P., Siess L., Lattanzio J.~C., Lau H.~H.~B., 2015, MNRAS, 446, 2599. 


\bibitem[\protect\citeauthoryear{Du et al.}{2016}]{Du2016} Du B., et al., 2016, ApJS, 227, 27 


\bibitem[\protect\citeauthoryear{Eisenstein et al.}{2006}]{Eisenstein2006} Eisenstein D.~J., et al., 2006, ApJS, 167, 40 

\bibitem[\protect\citeauthoryear{Farihi, G{\"a}nsicke \& Koester}{2013}]{Farihi2013} Farihi J., G{\"a}nsicke B.~T., Koester D., 2013, Sci, 342, 218

\bibitem[\protect\citeauthoryear{Fontaine, Brassard, \& Bergeron}{2001}]{Fontaine2001} Fontaine G., Brassard P., Bergeron P., 2001, PASP, 113, 409 

\bibitem[\protect\citeauthoryear{Gaia Collaboration, et al.}{2016}]{Gaia2016} Gaia Collaboration, et al., 2016, A\&A, 595, A1

\bibitem[\protect\citeauthoryear{Gaia Collaboration, et al.}{2018a}]{Gaia2018} Gaia Collaboration, et al., 2018, A\&A, 616, A1

\bibitem[\protect\citeauthoryear{Gaia Collaboration, et al.}{2018b}]{Gaia2018b} Gaia Collaboration, et al., 2018, A\&A, 616, A11

\bibitem[\protect\citeauthoryear{G{\"a}nsicke et al.}{2012}]{Gansicke2012} G{\"a}nsicke B.~T., Koester D., Farihi J., Girven J., Parsons S.~G., Breedt E., 2012, MNRAS, 424, 333


\bibitem[\protect\citeauthoryear{G{\"a}nsicke et al.}{2006}]{Gansicke2006} G{\"a}nsicke B.~T., Marsh T.~R., Southworth J., Rebassa-Mansergas A., 2006, Sci, 314, 1908

\bibitem[\protect\citeauthoryear{Gentile Fusillo et al.}{2015}]{Gentile2015} Gentile Fusillo N.~P., et al., 2015, MNRAS, 452, 765 


\bibitem[\protect\citeauthoryear{Gentile Fusillo et al.}{2018}]{Gentile2018} Gentile Fusillo N.~P., Tremblay P.-E., Jordan S., G{\"a}nsicke B.~T., Kalirai J.~S., Cummings J., 2018, MNRAS, 473, 3693 

\bibitem[\protect\citeauthoryear{Gentile Fusillo et al.}{2019}]{Gentile2019} Gentile Fusillo N.~P., et al., 2019, MNRAS, 482, 4570

\bibitem[\protect\citeauthoryear{Girven et al.}{2011}]{Girven2011} Girven J., G{\"a}nsicke B.~T., Steeghs D., Koester D., 2011, MNRAS, 417, 1210

\bibitem[\protect\citeauthoryear{Green, Schmidt, \& Liebert}{1986}]{Green1986} Green R.~F., Schmidt M., Liebert J., 1986, ApJS, 61, 305 



\bibitem[\protect\citeauthoryear{Guo et al.}{2019}]{Guo2019} Guo J.-C., et al., 2019, Research in Astronomy and Astrophysics, 19, 008


\bibitem[Guo et al.(2018)]{Guo2018} Guo, J.-C., Zhang, H.-W., Zhang, H.-H., et al.\ 2018, Research in Astronomy and Astrophysics, 18, 32.


\bibitem[\protect\citeauthoryear{Guo, Liu, \& Liu}{2016}]{Guo2016} Guo J.-C., Liu C., Liu J.-F., 2016, Research in Astronomy and Astrophysics, 16, 44 

\bibitem[\protect\citeauthoryear{Guo et al.}{2015a}]{Guo2015a} Guo J., Tziamtzis A., Wang Z., Liu J., Zhao J., Wang S., 2015, ApJ, 810, L17

\bibitem[\protect\citeauthoryear{Guo et al.}{2015b}]{Guo2015b} Guo J., Zhao J., Tziamtzis A., Liu J., Li L., Zhang Y., Hou Y., Wang Y., 2015, MNRAS, 454, 2787 


\bibitem[\protect\citeauthoryear{Han}{1998}]{Han1998} Han Z., 1998, MNRAS, 296, 1019 


\bibitem[\protect\citeauthoryear{Harris et al.}{2006}]{Harris2006} Harris H.~C., et al., 2006, AJ, 131, 571 


\bibitem[\protect\citeauthoryear{Heger et al.}{2003}]{Heger2003} Heger A., Fryer C.~L., Woosley S.~E., Langer N., Hartmann D.~H., 2003, ApJ, 591, 288 

\bibitem[\protect\citeauthoryear{Holberg et al.}{2016}]{Holberg2016} Holberg J.~B., Oswalt T.~D., Sion E.~M., McCook G.~P., 2016, MNRAS, 462, 2295

\bibitem[\protect\citeauthoryear{Hollands et al.}{2018}]{Hollands2018} Hollands M.~A., Tremblay P.-E., G{\"a}nsicke B.~T., Gentile-Fusillo N.~P., Toonen S., 2018, MNRAS, 480, 3942

\bibitem[\protect\citeauthoryear{Jiang et al.}{2013}]{Jiang2013} Jiang B., Luo A., Zhao Y., Wei P., 2013, MNRAS, 430, 986 

\bibitem[\protect\citeauthoryear{Jim{\'e}nez-Esteban et al.}{2018}]{Jimenez2018} Jim{\'e}nez-Esteban F.~M., Torres S., Rebassa-Mansergas A., Skorobogatov G., Solano E., Cantero C., Rodrigo C., 2018, MNRAS, 480, 4505

\bibitem[\protect\citeauthoryear{Kaiser et al.}{2010}]{Kaiser2010} Kaiser N., et al., 2010, SPIE, 7733, 77330E 


\bibitem[\protect\citeauthoryear{Kalirai et al.}{2009}]{Kalirai2009} Kalirai J.~S., Saul Davis D., Richer H.~B., Bergeron P., Catelan M., Hansen B.~M.~S., Rich R.~M., 2009, ApJ, 705, 408 


\bibitem[\protect\citeauthoryear{Kalirai et al.}{2008}]{Kalirai2008} Kalirai J.~S., Hansen B.~M.~S., Kelson D.~D., Reitzel D.~B., Rich R.~M., Richer H.~B., 2008, ApJ, 676, 594 


\bibitem[\protect\citeauthoryear{Kepler et al.}{2007}]{Kepler2007} Kepler S.~O., Kleinman S.~J., Nitta A., Koester D., Castanheira B.~G., Giovannini O., Costa A.~F.~M., Althaus L., 2007, MNRAS, 375, 1315 


\bibitem[\protect\citeauthoryear{Kepler et al.}{2015}]{Kepler2015} Kepler S.~O., et al., 2015, MNRAS, 446, 4078 


\bibitem[\protect\citeauthoryear{Kepler et al.}{2016}]{Kepler2016} Kepler S.~O., et al., 2016, MNRAS, 455, 3413 

\bibitem[\protect\citeauthoryear{Kepler et al.}{2019}]{Kepler2019} Kepler S.~O., et al., 2019, MNRAS, 486, 2169

\bibitem[\protect\citeauthoryear{Kilic et al.}{2019}]{Kilic2019} Kilic M., Bergeron P., Dame K., Hambly N.~C., Rowell N., Crawford C.~L., 2019, MNRAS, 482, 965

\bibitem[\protect\citeauthoryear{Kilic et al.}{2018}]{Kilic2018} Kilic M., Hambly N.~C., Bergeron P., Genest-Beaulieu C., Rowell N., 2018, MNRAS, 479, L113

\bibitem[\protect\citeauthoryear{Kilic et al.}{2017}]{Kilic2017} Kilic M., Munn J.~A., Harris H.~C., von Hippel T., Liebert J.~W., Williams K.~A., Jeffery E., DeGennaro S., 2017, ApJ, 837, 162 


\bibitem[\protect\citeauthoryear{Kilic, Stanek, \& Pinsonneault}{2007}]{Kilic2007} Kilic M., Stanek K.~Z., Pinsonneault M.~H., 2007, ApJ, 671, 761 


\bibitem[\protect\citeauthoryear{Kleinman et al.}{2004}]{Kleinman2004} Kleinman S.~J., et al., 2004, ApJ, 607, 426 


\bibitem[\protect\citeauthoryear{Kleinman et al.}{2013}]{Kleinman2013} Kleinman S.~J., et al., 2013, ApJS, 204, 5 


\bibitem[\protect\citeauthoryear{Koester}{2010}]{Koester2010} Koester D., 2010, MmSAI, 81, 921 


\bibitem[\protect\citeauthoryear{Koester et al.}{2001}]{Koester2001} Koester D., et al., 2001, A\&A, 378, 556 


\bibitem[\protect\citeauthoryear{Koester \& Kepler}{2015}]{Koester2015} Koester D., Kepler S.~O., 2015, A\&A, 583, A86


\bibitem[\protect\citeauthoryear{Kollmeier et al.}{2017}]{Kollmeier2017} Kollmeier J.~A., et al., 2017, arXiv, arXiv:1711.03234 

\bibitem[\protect\citeauthoryear{Kong \& Luo}{2019}]{Kong2019} Kong X., Luo A.-L., 2019, ASPC, 523, 91

\bibitem[\protect\citeauthoryear{Krzesinski et al.}{2009}]{Krzesinski2009} Krzesinski J., Kleinman S.~J., Nitta A., H{\"u}gelmeyer S., Dreizler S., Liebert J., Harris H., 2009, A\&A, 508, 339 


\bibitem[\protect\citeauthoryear{Lam et al.}{2019}]{Lam2019} Lam M.~C., et al., 2019, MNRAS, 482, 715

\bibitem[\protect\citeauthoryear{Li et al.}{2018}]{Li2018} Li Y.-B., et al., 2018, ApJS, 234, 31 


\bibitem[\protect\citeauthoryear{Liebert, Bergeron, \& Holberg}{2005}]{Liebert2005} Liebert J., Bergeron P., Holberg J.~B., 2005, ApJS, 156, 47 


\bibitem[\protect\citeauthoryear{Liu et al.}{2014}]{Liu2014} Liu C., et al., 2014, ApJ, 790, 110 

\bibitem[\protect\citeauthoryear{Liu et al.}{2019}]{Liu2019} Liu N., et al., 2019, RAA, 19, 75

\bibitem[\protect\citeauthoryear{Liu et al.}{2014}]{LiuXW2014} Liu X.-W. et al., 2014, in Feltzing S., Zhao G., Walton N. A., Whitelock P.,eds, IAU Symp. 298, Setting the Scene for Gaia and LAMOST, p. 310

\bibitem[\protect\citeauthoryear{Luo et al.}{2015}]{Luo2015} Luo A.-L., et al., 2015, RAA, 15, 1095 


\bibitem[\protect\citeauthoryear{Luri et al.}{2018}]{Luri2018} Luri X., et al., 2018, A\&A, 616, A9 


\bibitem[\protect\citeauthoryear{Manser et al.}{2019}]{Manser2019} Manser C.~J., et al., 2019, Sci, 364, 66


\bibitem[\protect\citeauthoryear{Markwardt}{2009}]{Markwardt2009} Markwardt C.~B., 2009, ASPC, 411, 251 


\bibitem[\protect\citeauthoryear{McCook \& Sion}{1999}]{McCook1999} McCook G.~P., Sion E.~M., 1999, ApJS, 121, 1 


\bibitem[\protect\citeauthoryear{McCook \& Sion}{1987}]{McCook1987} McCook G.~P., Sion E.~M., 1987, ApJS, 65, 603 


\bibitem[\protect\citeauthoryear{Melis et al.}{2012}]{Melis2012} Melis C., et al., 2012, ApJ, 751, L4

\bibitem[\protect\citeauthoryear{Munn et al.}{2017}]{Munn2017} Munn J.~A., et al., 2017, AJ, 153, 10

\bibitem[\protect\citeauthoryear{Pedregosa et al.}{2011}]{Pedregosa2011} Pedregosa F., et al., 2011, Journal of Machine Learning Research, 12, 2825

\bibitem[\protect\citeauthoryear{Rebassa-Mansergas et al.}{2015}]{Rebassa2015} Rebassa-Mansergas A., et al., 2015, MNRAS, 450, 743 


\bibitem[\protect\citeauthoryear{Rebassa-Mansergas et al.}{2011}]{Rebassa2011} Rebassa-Mansergas A., Nebot G{\'o}mez-Mor{\'a}n A., Schreiber M.~R., Girven J., G{\"a}nsicke B.~T., 2011, MNRAS, 413, 1121 

\bibitem[\protect\citeauthoryear{Ren et al.}{2018}]{Ren2018} Ren J.-J., Rebassa-Mansergas A., Parsons S.~G., Liu X.-W., Luo A.-L., Kong X., Zhang H.-T., 2018, MNRAS, 477, 4641

\bibitem[\protect\citeauthoryear{Rolland, Bergeron, \& Fontaine}{2018}]{Rolland2018} Rolland B., Bergeron P., Fontaine G., 2018, ApJ, 857, 56. doi:10.3847/1538-4357/aab713

\bibitem[\protect\citeauthoryear{Rowell}{2013}]{Rowell2013} Rowell N., 2013, MNRAS, 434, 1549 

\bibitem[\protect\citeauthoryear{Rowell \& Kilic}{2019}]{Rowell2019} Rowell N., Kilic M., 2019, MNRAS, 484, 3544


\bibitem[\protect\citeauthoryear{Tian et al.}{2017}]{Tian2017} Tian H.-J., et al., 2017, ApJS, 232, 4 


\bibitem[\protect\citeauthoryear{Toonen, Nelemans, \& Portegies Zwart}{2012}]{Toonen2012} Toonen S., Nelemans G., Portegies Zwart S., 2012, A\&A, 546, A70 

\bibitem[\protect\citeauthoryear{Torres et al.}{2019}]{Torres2019} Torres S., Cantero C., Rebassa-Mansergas A., Skorobogatov G., Jim{\'e}nez-Esteban F.~M., Solano E., 2019, MNRAS, 485, 5573


\bibitem[\protect\citeauthoryear{Tremblay \& Bergeron}{2009}]{Tremblay2009} Tremblay P.-E., Bergeron P., 2009, ApJ, 696, 1755 


\bibitem[\protect\citeauthoryear{Tremblay, Bergeron, \& Gianninas}{2011}]{Tremblay2011} Tremblay P.-E., Bergeron P., Gianninas A., 2011, ApJ, 730, 128 

\bibitem[\protect\citeauthoryear{Tremblay et al.}{2020}]{Tremblay2020} Tremblay P.-E., Hollands M.~A., Gentile Fusillo N.~P., McCleery J., Izquierdo P., G{\"a}nsicke B.~T., Cukanovaite E., et al., 2020, MNRAS, 497, 130. doi:10.1093/mnras/staa1892

\bibitem[\protect\citeauthoryear{Wei et al.}{2014}]{Wei2014} Wei P., Luo A., Li Y., Tu L., Wang F., Zhang J., Chen X., et al., 2014, AJ, 147, 101. doi:10.1088/0004-6256/147/5/101


\bibitem[\protect\citeauthoryear{Wood}{1995}]{Wood1995} Wood M.~A., 1995, LNP, 443, 41 


\bibitem[\protect\citeauthoryear{Wood}{1990}]{Wood1990} Wood M.~A., 1990, PhDT,  


\bibitem[\protect\citeauthoryear{York et al.}{2000}]{York2000} York D.~G., Adelman J., Anderson J.~E., Anderson S.~F., Annis J., Bahcall N.~A., Bakken J.~A., et al., 2000, AJ, 120, 1579. doi:10.1086/301513


\bibitem[\protect\citeauthoryear{Yuan et al.}{2015}]{Yuan2015} Yuan H.-B., et al., 2015, MNRAS, 448, 855 


\bibitem[\protect\citeauthoryear{Zhang et al.}{2013}]{Zhang2013} Zhang Y.-Y., et al., 2013, AJ, 146, 34 


\bibitem[\protect\citeauthoryear{Zhao et al.}{2012}]{Zhao2012} Zhao G., Zhao Y.-H., Chu Y.-Q., Jing Y.-P., Deng L.-C., 2012, RAA, 12, 723 


\bibitem[\protect\citeauthoryear{Zhao et al.}{2013}]{Zhao2013} Zhao J.~K., Luo A.~L., Oswalt T.~D., Zhao G., 2013, AJ, 145, 169 


\bibitem[\protect\citeauthoryear{Zhao et al.}{2012}]{Zhao2012} Zhao J.~K., Oswalt T.~D., Willson L.~A., Wang Q., Zhao G., 2012, ApJ, 746, 144 


\end{thebibliography}
\end{document}